\documentclass{article}

\usepackage{arxiv}

\usepackage[utf8]{inputenc} 
\usepackage[T1]{fontenc}    
\usepackage{hyperref}       
\usepackage{url}            
\usepackage{booktabs}       
\usepackage{amsfonts}       
\usepackage{nicefrac}       
\usepackage{tikz}
\usepackage{microtype}      
\usepackage{lipsum}
\usepackage{graphicx}
\usepackage{subcaption}
\usepackage{dsfont}
\usepackage{amsmath}
\usepackage[ruled,vlined]{algorithm2e}
\usepackage{xcolor}
\graphicspath{ {./images/} }
\newcommand{\Pf}{\mathbb{P}(\mathcal{F})}

\newcommand{\Pfpost}{\mathbb{P}(\mathcal{F}\mid\mathbf{y})}
\newcommand{\Pfpostest}{\hat{\mathbb{P}}(\mathcal{F}\mid\mathbf{y})}

\newcommand{\Qb}{\mathbf{Q}}
\newcommand{\Lb}{\mathbf{L}}

\title{Multiscale Structural Reliability Analysis in high dimensions with Tensor Trains and Physics-Augmented Neural Networks}

\author{
\textbf{Aryan Tyagi}$^{1}$ \qquad
\textbf{Alex de Beer}$^{2}$ \qquad
\textbf{Tiangang Cui}$^{2}$ \qquad
\textbf{Jan N. Fuhg}$^{1,3,\dagger}$ \\[1em]
\begin{minipage}{0.9\textwidth}
\raggedright
\small
$^{1}$Department of Aerospace Engineering \& Engineering Mechanics, The University of Texas at Austin, Austin, TX\\
$^{2}$School of Mathematics and Statistics, University of Sydney, New South Wales, Australia\\
$^{3}$The Oden Institute of Computational Science and Engineering, The University of Texas at Austin, Austin, TX\\
$^{\dagger}$Correspondence: \texttt{jan.fuhg@utexas.edu}
\end{minipage}
}

\begin{document}
\maketitle
\begin{abstract}
Structural reliability evaluation for composites constitutes a fundamentally high-dimensional multiscale problem, as microscale material uncertainties (e.g., fiber volume fraction and constituent moduli) must propagate to the macroscale and can be quantified as high-dimensional random fields. Conventional approaches are computationally intractable, as they rely on repeatedly solving tightly coupled partial differential equation systems across scales while contending with the exponential complexity inherent in high-dimensional uncertainty quantification (UQ).
This work introduces a scalable and physically consistent framework that addresses both bottlenecks simultaneously in the case of separation of scales and (anisotropic) linear elasticity. In particular, we couple a physics-augmented Voigt--Reuss Neural Network (VRNN) with the Deep Inverse Rosenblatt Transport (DIRT) method to estimate the posterior probability of multiscale structural failure. The VRNN is used to resolve the computationally expensive FE$^2$ scheme by providing a near-instantaneous evaluation of the homogenized stiffness tensor that is guaranteed to be symmetric, positive-definite, and strictly bounded within the Voigt--Reuss limits. This enables fast evaluation of the homogenized responses, which are required for the macroscale finite element model. The DIRT method constructs a sequence of functional tensor train approximations to efficiently store an approximation of the high-dimensional optimal importance sampling distribution for estimating the probability of failure under the VRNN surrogate. This mitigates the curse of dimensionality arising from the Karhunen--Loève (KL) expansion of the random fields. The framework is demonstrated on a three-dimensional heterogeneous benchmark problem, where the uncertainty in the microscale material properties is characterized by a Bayesian posterior distribution obtained from limited strain observations. Our results show that the DIRT-based estimator can provide low-variance estimates of failure probabilities in dimensions up to 150. This performance confirms that the combined VRNN-DIRT approach yields reliable failure probability estimates efficiently, providing a robust solution for scalable multiscale UQ in engineered composites.
\end{abstract}


\section{Introduction}
\label{sec:introduction}

Structural reliability analysis aims to quantify the safety of an engineering structure. This is typically achieved by modeling uncertain input parameters in a probabilistic framework and evaluating the failure domain through a Limit State Function (LSF) to estimate the probability of failure of the structure \cite{lsf1,lsf2,lsf3}. Over the past decades, a large number of numerical methods have been developed and used to efficiently obtain estimations of the probability of failure, such as Subset Simulation \cite{Subset_original}, Importance Sampling \cite{IS2,IS3}, and First and Second Order Reliability Methods (FORM and SORM) \cite{FORM_original, SORM_1}. In mechanical systems, uncertainties in the parameters can be aleatoric (inherent) or epistemic (arising from measurement errors). Most classical reliability methods assume probabilistic models are known a priori, and therefore implicitly treat uncertainty as aleatoric \cite{prior1,prior2,prior3,prior4,prior5}. However, recent efforts have also incorporated epistemic uncertainties through Bayesian updating of model parameters and uncertainties using noisy measurement data, leading to an updated (posterior) probability of failure \cite{DIRT_structural, BUS_KLE,BUS_varying, efficient_framework_Li2024}. This leads to the \textit{posterior} probability of failure, which reflects both inherent randomness and updated knowledge inferred from measurement data.
While probabilistic reliability methods quantify how uncertainties affect structural failure, a central question that remains is: \emph{where do these uncertainties actually originate and how should they be modeled in materials with complex internal architecture such as fiber-reinforced composites?}
Fiber-reinforced composite materials used in aerospace, mechanical, and civil engineering structures are often subjected to harsh operating conditions. In such composite materials, mesoscale fluctuations are present in the material properties of the structure \cite{uribe2021cross, BUS_varying}. However, the actual source of uncertainty is the microstructure of the composite. This can arise from manufacturing defects, variations in constituent moduli, or imperfections introduced during additive manufacturing.  These considerations naturally motivate modeling effective microstructural properties as random fields. 
This makes structural reliability for composite materials fundamentally a multiscale problem. The microscale uncertainties must propagate to the macroscale before influencing the structural response. Under the assumption of statistical and mechanical scale separation \cite{ostoja2006material,chen2024concurrent}, such multiscale problems can be solved using a hierarchical approach \cite{geers2003multiscale}. In complex geometries, this naturally involves solving a system of coupled partial differential equations (PDEs). Here, the associated boundary value problems (BVPs) at the microscale (called homogenization problems) are typically solved using Finite Element Methods (FEM) \cite{FE2_1,FE2_2,FE2_3,FE2_4}, or Fast Fourier Transform (FFT)-based solvers \cite{FFT_1, FFT_2}. Using these approaches, several studies have incorporated uncertainty in the geometric descriptors of the microscale unit cell \cite{imprecise, random_RVE}.  
However, computationally quantifying the effects of these uncertainties is generally computationally prohibitive, especially for spatially varying macrostructures. In particular, FE$^2$ (FE at macroscale -- FE at microscale) needs to solve a BVP at each integration point over the macroscale. While FE-FFT is faster than FE, increasingly fine voxel grids are required if the uncertainties arise due to small changes in the microstructure geometry (e.g., fiber radius). To address this, reduced-order and surrogate models for microstructural homogenization have been developed. These approaches replace the coupled PDE system with a macroscale PDE combined with a constitutive evaluation, similar to classical single-scale continuum mechanics. Historically, these reduced-order models were mostly based on intrusive projection methods based on proper orthogonal decomposition \cite{lenaerts2001proper}. However, parameterized homogenization, like that required for uncertainty quantification, is not straightforward with these methods \cite{radermacher2013comparison,radermacher2016pod}.
In the context of surrogate modeling, parameterization can be naturally included. Several machine learning architectures have therefore been applied, such as feedforward neural networks \cite{li2020multiscale,black2023deep}, graph neural networks \cite{vlassis2020geometric}, or Gaussian process regression \cite{fuhg2022local}.

However, the field has increasingly shifted towards the development of machine learning models that can intrinsically, i.e., by design, fulfill physical concepts and mechanistic assumptions instead of only being trained on data \cite{inverse_PANNs, ISNN, Kalina_2025,NN_homogenization_2,NN_homogenization_1}. In such architectures, fewer training labels are needed since these assumptions act as regularizers in the parametric space. Additionally, they have the potential to remain more accurate when extrapolating data \cite{fuhg2025review}. In the present work, we focus on fiber-reinforced composites with linearly elastic fiber and matrix materials. We rely on the recently introduced Voigt-Reuss Neural Networks (VRNNs) that structurally enforce that predicted stiffness matrices remain within the theoretical Voigt--Reuss bounds for a given microstructure \cite{keshav2025spectral}.

In this work, we assume that the microstructure is spatially variable over the macroscale (under the assumption of scale separation) and follows an unknown probability distribution, which can be inferred from limited macroscale observations through Bayesian updating. Even with efficient homogenization, reliability analysis remains computationally prohibitive due to the high-dimensional uncertainty in the microstructure field. Representing material properties as a random field is generally achieved by using a truncated Karhunen-Loeve (KL) expansion \cite{KLE_RF_1, voelsen2023pce, KLE_RF_2} where any realization of the field can be reconstructed by sampling standard Gaussian random variables. The KL expansion is generally truncated to capture a specific percentage of the energy in the field. As the number of KL modes increases, the dimensionality of the parameter space grows, making sampling exponentially more expensive (the \emph{curse of dimensionality} \cite{curtis2001prior}). In high-dimensional structural reliability problems, the failure region can become extremely concentrated in the parameter space, essentially necessitating the use of advanced sampling algorithms such as Importance Sampling.
This brings us to another bottleneck: selecting an effective biasing distribution for importance sampling. The optimal biasing distribution minimizes the variance of the estimator, but it depends on the unknown failure probability and the structure of the failure region, making it unavailable in practice \cite{bucklew2004introduction}. To address this challenge, we approximate the optimal biasing distribution using the Deep Inverse Rosenblatt Transport (DIRT) framework developed in Refs.~\cite{approxsampling_TT, DIRTRareEvent2024, DIRT2022} and used for structural reliability in Ref.~\cite{DIRT_structural}. The Tensor Train (TT) decomposition enables an efficient representation of this distribution, with storage costs that scale linearly with dimension. Without low-rank tensor formats, storing or sampling the optimal biasing distribution is infeasible due to exponential memory growth. The DIRT method builds this distribution via the Inverse Rosenblatt Transformation (IRT) in TT format. The two bottlenecks discussed above are not separate issues, but are actually related: treating the microstructure descriptors as random fields leads to a large number of KL modes, increasing the dimensionality of the sampling problem and leading to a need for scalable and fast homogenization. Thus, scalable reliability analysis for heterogeneous composite materials requires addressing both bottlenecks simultaneously. The main contribution of this work is a multiscale reliability framework that combines VRNN-based physics-constrained homogenization with DIRT-based transport sampling, enabling scalable reliability analysis for heterogeneous composite materials with high-dimensional random-field uncertainty.

The remainder of the paper is organized as follows. Section~\ref{sec:theory} introduces the theoretical framework and methods used in this study, including structural reliability analysis, importance sampling, the VRNN surrogate model, and the DIRT methodology. Section~\ref{sec:example} presents the high-dimensional numerical experiment used to evaluate the proposed framework. Finally, Section~\ref{sec:conclusion} summarizes the main findings and discusses limitations and directions for future research.

\section{Theory}
\label{sec:theory}



\subsection{Multiscale Modeling}
\label{sec:multiscale}

Engineering materials such as fiber-reinforced composites exhibit mechanical behavior governed by features spanning multiple spatial scales. While structural components operate at the macroscale, their effective material response is dictated by microstructural characteristics such as fiber orientation, geometry, volume fraction, and constituent properties. Multiscale modeling provides a framework for linking these scales and for propagating microscale uncertainty to macroscale predictions. Because explicitly resolving all microstructural features in a macroscale mesh is computationally infeasible, homogenization approaches are commonly used to replace the detailed microstructure with an effective constitutive description. In this setting, a representative microstructural sample, often referred to as a Representative Volume Element (RVE), is used to compute the homogenized constitutive response associated with a given set of microstructural parameters.
An RVE is a sufficiently large sample of the microstructure that captures its statistical features while remaining small relative to the structural scale so that it can be associated with a material point at the macroscale. The applicability of RVE-based homogenization relies on the \emph{separation of scales} assumption
\begin{equation}
    \varepsilon = \frac{l}{L} \ll 1 ,
\end{equation}
where $l$ is a characteristic microscale length (e.g., fiber diameter) and $L$ is the characteristic macroscale length (e.g., structural dimension). When this condition is satisfied, the macroscopic strain field may be regarded as uniform over the RVE, and the heterogeneous composite can be replaced by an equivalent homogeneous material characterized by an effective stiffness tensor $\mathbb{C}^{\textrm{hom}}$.
Given any microscopic field $(\bullet)$, its volume average over the RVE is defined as
\begin{equation}
\langle (\bullet) \rangle :=
\frac{1}{V_{\mathrm{RVE}}}
\int_{\mathcal{B}^{\mathrm{RVE}}} (\bullet)\, dV .
\end{equation}

Let $\boldsymbol{\varepsilon}(\mathbf{x})$ denote the microscopic small–strain tensor
and $\boldsymbol{\sigma}(\mathbf{x})$ the corresponding Cauchy stress. The macroscopic
strain and stress are then introduced through
\begin{equation}    
\bar{\boldsymbol{\varepsilon}} = \langle \boldsymbol{\varepsilon} \rangle, \quad 
\bar{\boldsymbol{\sigma}} = \langle \boldsymbol{\sigma} \rangle ,
\end{equation}

and homogenization seeks $\mathbb{C}^{\textrm{hom}}$ such that
\begin{equation}
\bar{\boldsymbol{\sigma}} = \mathbb{C}^{\textrm{hom}}:\bar{\boldsymbol{\varepsilon}} .
\end{equation}

In classical computational homogenization, the constitutive parameters of each RVE (e.g.,~fiber volume fraction $v_f$, fiber modulus $E_f$, and matrix modulus $E_m$) are taken as fixed. In the present work, these quantities vary spatially across the macroscale domain and are modeled as correlated random fields. Consequently, each macroscopic integration point corresponds to a distinct local microstructure and, therefore, a distinct effective stiffness tensor $\mathbb{C}^{\textrm{hom}}$.

\paragraph{On scale separation for spatially varying microstructure fields}
Introducing spatially varying random fields for quantities such as fiber radius, modulus, or volume fraction raises the question of whether classical scale separation still applies. In the present setting, the random field varies over the macroscale, while each RVE still contains many microscopic features whose characteristic length is much smaller than the macroscale length. A convenient way to formalize this is through a slow–fast decomposition of a generic microstructure descriptor \cite{Allaire_1992}
\begin{equation}
    t(x) = t_0(x) + t_1\!\left(\frac{x}{\varepsilon}\right),
\label{eq:slow_fast}
\end{equation}
where $t_0(x)$ represents a slowly varying trend across the structure (e.g., a long-range correlated drift in fiber radius), $t_1(x/\varepsilon)$ captures the rapid microscale oscillations within each RVE, and $\varepsilon=\ell/L\ll 1$ is the ratio of micro- to macro-length scales. Under this structure, long-range correlation in $t_0(x)$ does not induce mechanical interaction between neighboring RVEs; the RVE problem at each point is still driven solely by the local macroscopic strain field. Consequently, the homogenized stiffness tensor remains well defined but becomes spatially dependent through $t_0(x)$. This is a standard situation in heterogeneous materials with slowly varying coefficients and is fully compatible with first-order computational homogenization. 

A direct FE$^2$ approach would require solving a separate RVE problem at every quadrature point for each realization of the microstructural random fields, which is computationally prohibitive for uncertainty quantification. In this work, we replace these repeated homogenization solves with the VRNN surrogate introduced in Section~\ref{sec:vrnn}, which provides instantaneous evaluations of the homogenized stiffness tensor $\mathbb{C}_{\mathrm{eff}}$ conditioned on the local microstructural parameters.
The microscopic and macroscopic fields must satisfy the Hill--Mandel macro-homogeneity condition
\begin{equation}
\langle \boldsymbol{\sigma} : \boldsymbol{\varepsilon} \rangle
=
\bar{\boldsymbol{\sigma}} : \bar{\boldsymbol{\varepsilon}},
\end{equation}
which states that the macroscopic stress power equals the volume-averaged microscopic stress power. This requirement guarantees energetic equivalence between the heterogeneous microstructure and its homogeneous effective representation.
A class of boundary conditions can be used to satisfy the Hill--Mandel condition, including uniform displacement, uniform traction, and periodic boundary conditions. In the present work, periodic boundary conditions are adopted because they satisfy the Hill--Mandel condition exactly and provide effective responses that typically lie between the Voigt and Reuss bounds. This assumption is commonly used in computational homogenization and remains appropriate when microstructural parameters vary slowly across the macroscale \cite{Allaire_1992}. Under periodic boundary conditions,
\begin{equation}
\mathbf{u}^{+} - \mathbf{u}^{-}
= \bar{\boldsymbol{\varepsilon}}(\mathbf{x}^{+}-\mathbf{x}^{-}),\quad
\boldsymbol{\sigma}^{+}\mathbf{n}^{+}
= -\boldsymbol{\sigma}^{-}\mathbf{n}^{-},
\end{equation}
where $(\cdot)^{\pm}$ denote pairs of opposite boundary points.
To compute $\mathbb{C}_{\mathrm{eff}}$, six linearly independent macroscopic strain states $\{\bar{\boldsymbol{\varepsilon}}^{(k)}\}_{k=1}^6$ are prescribed on the RVE boundary. For each loading case, the corresponding microscopic equilibrium problem,
\begin{equation}
\nabla \cdot \boldsymbol{\sigma}^{(k)} = \mathbf{0},\boldsymbol{\sigma}^{(k)} = \mathbb{C}(\mathbf{x}) :
\boldsymbol{\varepsilon}^{(k)},
\end{equation}
is solved under the chosen boundary conditions, and the homogenized stress is obtained as
\begin{equation}
\bar{\boldsymbol{\sigma}}^{(k)} = \langle \boldsymbol{\sigma}^{(k)} \rangle.
\end{equation}
The $6\times 6$ effective stiffness matrix is then assembled by matching
\begin{equation}
\bar{\boldsymbol{\sigma}}^{(k)}
=
\mathbb{C}_{\mathrm{eff}}
:
\bar{\boldsymbol{\varepsilon}}^{(k)},
k = 1,\ldots,6 .
\end{equation}

\subsection{Voigt--Reuss Neural Networks}
\label{sec:vrnn}

\begin{figure}
    \centering
    \includegraphics[width=0.75\linewidth]{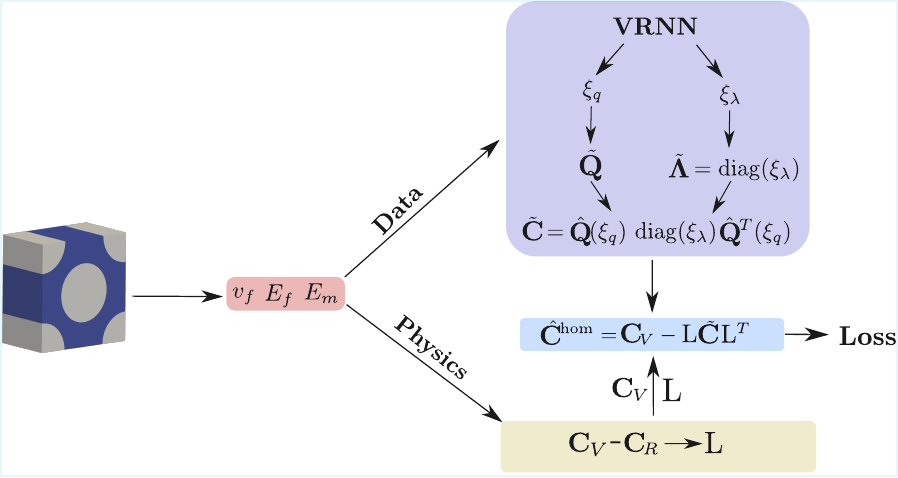}
    \caption{Workflow of the VRNN used for homogenization. 
The input RVE provides microscale descriptors $(v_f, E_f, E_m)$, which are passed through the VRNN 
to produce normalized spectral parameters: orthogonal matrix parameters $\xi_q$ and eigenvalues 
$\xi_\lambda$. These are converted into an orthogonal matrix $\hat{\Qb}$ and a diagonal eigenvalue matrix 
$\tilde{\mathbf{\Lambda}}$, yielding the normalized tensor 
$\tilde{\mathbf{C}} = \hat{\Qb}\tilde{\mathbf{\Lambda}}\hat{\Qb}^\mathsf{T}$. 
A physics-augmented reconstruction maps this tensor to the homogenized stiffness 
$\hat{\mathbf{C}}^{\mathrm{hom}} = \mathbf{C}_V - \Lb\,\tilde{\mathbf{C}}\,\Lb^\mathsf{T}$, guaranteeing symmetry, positive 
semi-definiteness, and adherence to the Voigt--Reuss bounds $\mathbf{C}_R \preceq \hat{\mathbf{C}}^{\mathrm{hom}} \preceq \mathbf{C}_V$.}
\label{fig:placeholder}
\end{figure}

Classical FE$^2$ homogenization becomes prohibitive when required inside Monte Carlo sampling, importance sampling, or Bayesian inference frameworks, where a large number of forward evaluations may be necessary. This motivates the use of a physically consistent surrogate model for RVE homogenization. The Voigt--Reuss Neural Network (VRNN) surrogate introduced in Ref.~ \cite{keshav2025spectral} learns the RVE homogenization map. In this work, we restrict the input parameters to fiber volume fraction $v_f$, fiber Young's modulus $E_f$, and matrix Young's modulus $E_m$. The homogenization mapping is thus expressed as follows
\begin{equation}
    \chi = (\mathrm{v_f, E_f, E_m}) \mapsto \mathbb{C}^{\mathrm{hom}} \ .
\end{equation}
VRNN provides a physics-augmented surrogate model for mapping microstructural and material parameters to effective constitutive tensors while guaranteeing strict physical admissibility by enforcing physical constraints such as symmetry, positive definiteness, and Voigt--Reuss bounds. The VRNN achieves this by embedding the analytical Voigt and Reuss limits of heterogeneous media directly into the learning process. 


In order to formalize the spectral normalization, it is assumed that the 4$^{th}$-order homogenized stiffness tensor $\mathbb{C}^{\text{hom}}$ can be expressed as a matrix $\mathbf{C}^{\text{hom}}\in\mathbb{R}^{m\times m}$ and that the upper and lower bounds $\mathbf{C}_V$ and $\mathbf{C}_R$, respectively, exist.
For a heterogeneous composite with local constitutive tensor $\mathbf{C}(\mathbf{x})$, the homogenized stiffness tensor $\mathbf{C}^{\text{hom}}$ satisfies the strict Löwner bounds
\begin{equation}
\mathbf{C}_{\text{R}} \preceq \mathbf{C}^{\text{hom}} \preceq \mathbf{C}_{\text{V}} \ ,
\end{equation}
where $\mathbf{C}_{\text{V}}$ and $\mathbf{C}_{\text{R}}$ denote the classical Voigt (uniform strain) and Reuss (uniform stress) bounds, respectively. These bounds can be generalized to any symmetric positive definite tensor field, ensuring that the predicted effective response lies within a physically meaningful set.
This can be rewritten as 
\begin{equation}
\label{eq:lowner}
 \mathbf{0} \preceq \mathbf{C}_{\text{V}} - \mathbf{C}^{\text{hom}} \preceq \mathbf{C}_{\text{V}} - \mathbf{C}_{\text{R}} \ .
\end{equation}
The right side of the above equation is positive-semidefinite and symmetric by definition, so it can be diagonalized as follows
\begin{equation}
    \mathbf{C}_{\text{V}} - \mathbf{C}_{\text{R}} +\mathcal{\mathbf{O}}(\epsilon)= \tilde{\mathbf{Q}}\tilde{\mathbf{\Lambda}}\tilde{\mathbf{Q}}
\end{equation}
where $\tilde{\mathbf{Q}}$ is an orthogonal matrix, $\tilde{\mathbf{\Lambda}} = \mathrm{diag}(\xi_{\lambda_1},\dots,\xi_{\lambda_m})$ is a diagonal matrix of normalized eigenvalues constrained to $[0,1]$ and $\epsilon$ is a numerical tolerance parameter that ensures $\mathbf{\Lambda}$ remains positive definite. Next, we can obtain $L$ as
\begin{equation}
    \Lb = \tilde{\Qb}\sqrt{\tilde{\mathbf{\Lambda}}}
\end{equation}
from the factorization of $(\mathbf{C}_{\text{V}} - \mathbf{C}_{\text{R}})$ and its pseudo-inverse $L^+$ as
\begin{equation}
    \Lb^+ = \sqrt{\tilde{\mathbf{\Lambda}}^\mathrm{-1}}\tilde{\Qb}^\mathsf{T} \ .     
\end{equation}
Then, \eqref{eq:lowner} can be reformulated to obtain normalized bounds between 0 and $I$ as 
\begin{equation}
    \mathbf{0} \preceq \tilde{\mathbf{C}} = \mathbf{L}^+(\mathbf{C}_{\text{V}} - \mathbf{C}^{\text{hom}})\mathbf{L}^{+\textsf{T}} \preceq \mathbf{L}^+(\mathbf{C}_{\text{V}} - \mathbf{C}_{\text{R}})\mathbf{L}^{+\textsf{T}} = \mathbf{I} + \mathcal{\mathbf{O}}(\epsilon) \ .
\end{equation}
The idea behind VRNN is to rely on the normalized spectral coefficients $\tilde{\mathbf{C}}$ as network output instead of training directly to $\mathbf{C}^{\text{hom}}$, which guarantees that predictions always remain within the Voigt--Reuss bounds by construction.
Following Ref.~\cite{keshav2025spectral}, the VRNN employs a fully connected feedforward neural network that maps microstructural descriptors and constituent material parameters $\boldsymbol{\chi}$ to the normalized spectral representation $\tilde{\mathbf{C}}$. Each sample provides inputs $\boldsymbol{\chi} = [v_f, \, E_f, \, E_m]$, where $v_f$ denotes the fiber volume fraction, and $E_f$, $E_m$ are the Young's moduli of the fiber and matrix phases, respectively. The network predicts both the orthogonal matrix parameters $\xi_q \in \mathbb{R}^{m(m-1)/2}$ and the eigenvalues $\xi_\lambda \in [0,1]^m$, which together define $\tilde{\mathbf{C}}$:
\begin{equation}
    \tilde{\mathbf{C}} = \hat{\Qb}(\xi_q) \, \mathrm{diag}(\xi_\lambda) \, \hat{\Qb}^\mathrm{T}(\xi_q).
\end{equation}
The final effective stiffness tensor is recovered through the inverse spectral transformation:
\begin{equation}
\hat{{\mathbf{C}}}^{\text{hom}} = {\mathbf{C}}_{\text{V}} - \Lb \, \tilde{\mathbf{C}} \, \Lb^\mathrm{T}.
\end{equation}
The loss function measures the normalized Frobenius distance between the predicted and reference normalized spectral tensors reconstructed from the eigenvalue–eigenvector parameterization
\begin{equation}
\mathcal{L} = 
\frac{1}{\sqrt{m}}
\, \mathbb{E}
\left[
\left\|
\tilde{\mathbf{C}}^{\text{pred}} -
\tilde{\mathbf{C}}^{\text{true}}
\right\|_F
\right],
\end{equation}
where each normalized tensor $\tilde{\mathbf{C}} = Q \, \mathrm{diag}(\boldsymbol{\xi}_\lambda) \, Q^\mathrm{T}$ is reconstructed from the learned eigenvalues $\boldsymbol{\xi}_\lambda$ and orthogonal parameters $Q$ recovered from the network outputs. This formulation enforces rotation-consistent comparisons between full symmetric tensors, rather than elementwise eigenvalue errors, and provides a dimensionless and scale-invariant measure of error bounded in $[0,1]$.

\subsection{Structural Reliability Analysis}
\label{sec:reliability}
In structural reliability analysis and structural health monitoring, the performance of an engineering system is expressed as a function of uncertain parameters such as material properties, external loads, or operational conditions. We denote these parameters by $\boldsymbol{\theta} \in \boldsymbol{\Theta} \subseteq \mathbb{R}^{d}$. In this work, uncertainties enter through the microscale descriptors 
$\{v_f(\mathbf{x}), E_f(\mathbf{x}), E_m(\mathbf{x})\}$ that govern the macroscale behavior. Each random field is represented using a KL expansion,
\begin{equation}
    \boldsymbol{\theta}(\mathbf{x}, \omega) 
    = \bar{\boldsymbol{\theta}}(\mathbf{x}) 
    + \sum_{i=1}^{d} \sqrt{\lambda_i}\, \boldsymbol{\phi}_i(\mathbf{x})\, \xi_i(\omega),
\end{equation}
where $\bar{\boldsymbol{\theta}}$ denotes the mean field, $\{(\lambda_i,\boldsymbol{\phi}_i)\}_{i=1}^{d}$ denote the eigenpairs of the field covariance kernel, and the KL coefficients $\boldsymbol{\xi} = [\xi_1, \ldots, \xi_d]$ are independent standard normal random variables. The system performance is represented by a performance function $g : \boldsymbol{\Theta} \rightarrow \mathbb{R}$; in particular, the system is said to fail if $g(\boldsymbol{\theta}) < 0$, and the corresponding set of parameters defines the failure region, $\mathcal{F} := \{\boldsymbol{\theta} \in \boldsymbol{\Theta} : g(\boldsymbol{\theta})<0\}$. 

Our aim is to estimate the probability of the system failing, subject to the uncertainty in the parameters. Following Ref.~\cite{DIRTRareEvent2024}, we adopt a Bayesian perspective when characterizing the uncertainty in the parameters, and consider two possible settings. The first is the \emph{a priori} setting, in which the system parameters follow a tractable \emph{prior} density $\pi_{\mathrm{pr}}$, with a known normalizing constant. In many applications, $\pi_{\mathrm{pr}}$ is chosen to be a multivariate Gaussian density. The corresponding \emph{prior} probability of failure is defined as
\begin{equation}
    \Pf := \int_{\Theta}\mathbb{I}_{\mathcal{F}}(\boldsymbol{\theta})\,\pi_{\text{pr}}(\boldsymbol{\theta})\, \text{d}\boldsymbol{\theta} = \mathbb{E}_{\pi_{\text{pr}}}[\mathbb{I}_{\mathcal{F}}(\boldsymbol{\theta})], 
    \label{eq:prior_pf}
\end{equation}
where $\mathbb{I}_{\mathcal{F}}: \boldsymbol{\Theta} \rightarrow \{0,1\}$ is the indicator function associated with the failure region $\mathcal{F}$.

The second setting is the \emph{a posteriori} setting, in which we have access to data (for instance, measurements of strains or displacements) which can be used to update the prior density to a posterior density, $\pi^{y}$, of the form
\begin{equation}
    \pi^{y}(\boldsymbol{\theta}) = \frac{1}{Z}\mathcal{L}^{y}(\boldsymbol{\theta})\pi_{\mathrm{pr}}(\boldsymbol{\theta}), \qquad Z := \int_{\boldsymbol{\Theta}}\mathcal{L}^{y}(\boldsymbol{\theta})\pi_{\mathrm{pr}}(\boldsymbol{\theta})\,\mathrm{d}\boldsymbol{\theta}. \label{eq:posterior}
\end{equation}
In the above, $\mathcal{L}^{y}$ denotes the likelihood function, which encodes the probability of the data being observed under a particular realization of the parameters, and $Z$ denotes an unknown normalizing constant which acts to ensure the posterior is a valid probability density. The corresponding posterior failure probability is defined as
\begin{equation}
    \Pfpost := \frac{1}{Z}\int_{\boldsymbol{\Theta}}\mathbb{I}_{\mathcal{F}}(\boldsymbol{\theta})\mathcal{L}^{y}(\boldsymbol{\theta})\pi_{\mathrm{pr}}(\boldsymbol{\theta})\, \mathrm{d}\boldsymbol{\theta}.
    \label{eq:post_pf}
\end{equation}
In the applications that motivate this work, the failure probabilities of interest are generally on the order of $10^{-3}$ or less \cite{beck2015rareeventsimulation}, rendering direct sampling approaches such as Monte Carlo sampling (MCS) or Markov chain Monte Carlo (MCMC) inefficient. To illustrate this point, it is instructive to introduce a standard Monte Carlo estimator of the prior probability of failure, which takes the form
\begin{equation}
    \Pf \approx \hat{\mathbb{P}}(\mathcal{F}) = \frac{1}{N}\sum_{i=1}^{N}\mathbb{I}_{\mathcal{F}}(\boldsymbol{\theta}^{(i)}), \qquad \{\boldsymbol{\theta}^{(i)}\}_{i=1}^{N} \sim \pi_{\mathrm{pr}}(\boldsymbol{\theta}). \label{eq:mc_estimate}
\end{equation}
The coefficient of variation (CoV) of estimate \eqref{eq:mc_estimate}, which provides a measure of accuracy, is defined as the ratio between the standard deviation and expectation of the estimate; that is,
\begin{equation}
    \text{CoV} := \frac{\sqrt{\mathbb{V}_{\pi_{\mathrm{pr}}}[\hat{\mathbb{P}}_{\mathcal{F}}]}}{\mathbb{E}_{\pi_{\mathrm{pr}}}[\hat{\mathbb{P}}_{\mathcal{F}}]} = \sqrt{\frac{1-\Pf}{N\Pf}},
\end{equation}
where $N$ denotes the total number of samples used to compute the estimate. If, for example, $\Pf = 10^{-5}$ and the target coefficient of variation of Eq.~\eqref{eq:mc_estimate} is $10\%$, approximately $N=10^{7}$ samples are required. This is prohibitively large when each sample involves a finite element simulation.

These challenges motivate the use of \emph{importance sampling} (IS), which aims to reduce estimator variance by biasing the sampling distribution towards the failure region, and correcting for this bias through a reweighting procedure. The following section briefly reviews the importance sampling approach used in this work.

\subsection{Importance Sampling} \label{sec:IS}

To introduce the idea of importance sampling for failure probability estimation, we begin by considering the \emph{a priori} setting. First, note that the prior probability of failure (Eq.~\ref{eq:prior_pf}) can be recast as the expectation
\begin{equation}
\begin{aligned}
        \Pf &= \mathbb{E}_{\pi_{\mathrm{pr}}}[\mathbb{I}_{\mathcal{F}}(\boldsymbol{\theta})] \\
    &= \mathbb{E}_{p}[\mathbb{I}_{\mathcal{F}}(\boldsymbol{\theta})\pi_{\mathrm{pr}}(\boldsymbol{\theta}) / p(\boldsymbol{\theta})],
\end{aligned}
\end{equation}
where $p$ denotes a \emph{biasing density} chosen such that $\mathrm{supp}(\mathbb{I}_{\mathcal{F}}\pi_{\mathrm{pr}}) \subseteq \mathrm{supp}(p)$. The corresponding importance sampling estimator for the prior probability of failure is defined as 
\begin{equation}
    \mathbb{P}(\mathcal{F}) \approx \hat{\mathbb{P}}(\mathcal{F}) := \frac{1}{N}\sum_{i=1}^{N} w^{(i)}\mathbb{I}_{\mathcal{F}}(\boldsymbol{\theta}^{(i)}), \qquad w^{(i)} := \frac{\pi_{\mathrm{pr}}(\boldsymbol{\theta}^{(i)})}{\pi_{\mathrm{bias}}(\boldsymbol{\theta}^{(i)})},
\end{equation}
where $\{\boldsymbol{\theta}^{(i)}\}_{i=1}^{N} \sim p(\boldsymbol{\theta})$. This estimator is unbiased, and its variance can be significantly reduced by selecting an appropriate biasing density. In particular, the optimal biasing density, $p^{*}$, takes the form
\begin{equation}
    p^{*}(\boldsymbol{\theta}) := \frac{\mathbb{I}_{\mathcal{F}}(\boldsymbol{\theta})\pi_{\mathrm{pr}}(\boldsymbol{\theta})}{\int_{\boldsymbol{\Theta}}\mathbb{I}_{\mathcal{F}}(\boldsymbol{\theta}')\pi_{\mathrm{pr}}(\boldsymbol{\theta}') \, \mathrm{d}\boldsymbol{\theta}'},
\end{equation}
and results in an estimator with zero variance \cite{intro_rare, mcbook}. The optimal biasing density cannot be used directly because of the unknown normalizing constant (which is in fact the target failure probability); however, it provides an indication as to the form of biasing densities that will produce estimators with low variance.

We now consider the \emph{a posteriori} setting, which is slightly more complex as a result of the unknown normalizing constant of the posterior. First, note that the posterior probability of failure (Eq.~\ref{eq:post_pf}) can be recast as
\begin{equation}
    \begin{aligned}
    \Pfpost &= \frac{\mathbb{E}_{\mathrm{pr}}[\mathbb{I}_{\mathcal{F}}(\boldsymbol{\theta})\mathcal{L}^{y}(\boldsymbol{\theta})]}{\mathbb{E}_{\mathrm{pr}}[\mathcal{L}^{y}(\boldsymbol{\theta})]} \label{eq:is_aposteriori} \\
    &= \frac{\mathbb{E}_{p^{y}}[\mathbb{I}_{\mathcal{F}}(\boldsymbol{\theta})\mathcal{L}^{y}(\boldsymbol{\theta})\pi_{\mathrm{pr}}(\boldsymbol{\theta})/p^{y}(\boldsymbol{\theta})]}{\mathbb{E}_{q^{y}}[\mathcal{L}^{y}(\boldsymbol{\theta})\pi_{\mathrm{pr}}(\boldsymbol{\theta})/q^{y}(\boldsymbol{\theta})]},        
    \end{aligned}
\end{equation}
where $p^{y}$ and $q^{y}$ are biasing densities chosen such that $\mathrm{supp}(\mathbb{I}_{\mathcal{F}}\mathcal{L}^{y}\pi_{\mathrm{pr}}) \subseteq \mathrm{supp}(p^{y})$ and $\mathrm{supp}(\mathcal{L}^{y}\pi_{\mathrm{pr}}) \subseteq \mathrm{supp}(q^{y})$. An importance sampling estimator for the numerator of Eq.~\eqref{eq:is_aposteriori} is given by 
\begin{equation}
    \mathbb{E}_{\mathrm{pr}}[\mathbb{I}_{\mathcal{F}}(\boldsymbol{\theta})\mathcal{L}^{y}(\boldsymbol{\theta})] \approx \hat{P}^{y} := \frac{1}{N}\sum_{i=1}^{N}w^{(i)} \mathbb{I}_{\mathcal{F}}(\boldsymbol{\theta}^{(i)})\mathcal{L}^{y}(\boldsymbol{\theta}^{(i)}), \qquad w^{(i)} := \frac{\pi_{\mathrm{pr}}(\boldsymbol{\theta}^{(i)})}{p^{y}(\boldsymbol{\theta}^{(i)})},
\end{equation}
where $\{\boldsymbol{\theta}^{(i)}\}_{i=1}^{N} \sim p^{y}(\boldsymbol{\theta})$. Similarly, an importance sampling estimator for the denominator of Eq.~\eqref{eq:is_aposteriori} is given by 
\begin{equation}
    \mathbb{E}_{\mathrm{pr}}[\mathcal{L}^{y}(\boldsymbol{\theta})] \approx \hat{Q}^{y} := \frac{1}{N}\sum_{i=1}^{N}w^{(i)} \mathcal{L}^{y}(\boldsymbol{\theta}^{(i)}), \qquad w^{(i)} := \frac{\pi_{\mathrm{pr}}(\boldsymbol{\theta}^{(i)})}{q^{y}(\boldsymbol{\theta}^{(i)})},
\end{equation}
where $\{\boldsymbol{\theta}^{(i)}\}_{i=1}^{N} \sim q^{y}(\boldsymbol{\theta})$. The corresponding importance sampling estimator for the posterior probability takes the form
\begin{equation}
    \Pfpost \approx \Pfpostest := \frac{\hat{P}^{y}}{\hat{Q}^{y}}. \label{eq:is_post_est}
\end{equation}
We note that the estimator Eq.~\eqref{eq:is_post_est} is biased (it is a ratio of two unbiased estimators); however, as shown in Ref.~\cite{DIRTRareEvent2024}, this bias is small. The optimal choices for densities $p^{y}$ and $q^{y}$, denoted by $p^{y*}$ and $q^{y*}$ respectively, are
\begin{equation}
    p^{y*}(\boldsymbol{\theta}) := \frac{\mathbb{I}_{\mathcal{F}}(\boldsymbol{\theta})\mathcal{L}^{y}(\boldsymbol{\theta})\pi_{\mathrm{pr}}(\boldsymbol{\theta})}{\int_{\boldsymbol{\Theta}}\mathbb{I}_{\mathcal{F}}(\boldsymbol{\theta}')\mathcal{L}^{y}(\boldsymbol{\theta}')\pi_{\mathrm{pr}}(\boldsymbol{\theta}') \, \mathrm{d}\boldsymbol{\theta}'}, \qquad 
    q^{y*}(\boldsymbol{\theta}) := \frac{\mathcal{L}^{y}(\boldsymbol{\theta})\pi_{\mathrm{pr}}(\boldsymbol{\theta})}{\int_{\boldsymbol{\Theta}}\mathcal{L}^{y}(\boldsymbol{\theta}')\pi_{\mathrm{pr}}(\boldsymbol{\theta}') \, \mathrm{d}\boldsymbol{\theta}'}. \label{eq:optimal_biasing_post}
\end{equation}
It is useful to note that the optimal biasing density $q^{y*}$ coincides with the posterior density (Eq.~\ref{eq:posterior}).

\subsection{Deep Inverse Rosenblatt Transport} \label{sec:dirt}

To carry out importance sampling efficiently, we use the deep importance sampling framework developed in Ref.~\cite{DIRTRareEvent2024}, which builds an approximation to the optimal biasing density $p^{*}$ (in the \emph{a priori} setting), or the optimal biasing densities $p^{y*}$ and $q^{y*}$ (in the \emph{a posteriori} setting) using the DIRT algorithm \cite{approxsampling_TT, DIRT2022}. Here, we provide a brief overview of the DIRT algorithm, which constructs a deterministic mapping between a tractable reference density and an approximation to the target density. For additional details, see \cite{approxsampling_TT, DIRTRareEvent2024, DIRT2022}. Central to the DIRT algorithm is the concept of pushforward and pullback densities (see, e.g., \cite{ramgraber2025friendly}). In particular, given random variables $\boldsymbol{u}$ and $\boldsymbol{\theta}$, with densities $\rho$ and $\pi$ respectively, and a diffeomorphism (i.e., a differentiable function with a differentiable inverse) $\mathcal{S}$ such that $\boldsymbol{\theta} = \mathcal{S}(\boldsymbol{u})$, the pushforward density of $\rho$ under $\mathcal{S}$, denoted by $\mathcal{S}_{\sharp}\rho$, takes the form
\begin{equation}
    \mathcal{S}_{\sharp}\rho(\boldsymbol{\theta}) = \rho(\mathcal{S}^{-1}(\boldsymbol{\theta}))|\nabla_{\boldsymbol{u}}\mathcal{S}^{-1}(\boldsymbol{\theta})| = \pi(\boldsymbol{\theta}).
\end{equation}
Similarly, the pullback density of $\pi$ under $\mathcal{S}$, denoted by $\mathcal{S}^{\sharp}\pi$, takes the form
\begin{equation}
    \mathcal{S}^{\sharp}\pi(\boldsymbol{u}) = \pi(\mathcal{S}(\boldsymbol{u}))|\nabla_{\boldsymbol{u}}\mathcal{S}(\boldsymbol{u})| = \rho(\boldsymbol{u}).
\end{equation}

The DIRT algorithm approximates an arbitrary probability density function using a layered composition of functional tensor train (FTT) representations \cite{bigoni2016spectral, gorodetsky2019continuous}. To construct an FTT approximation to an arbitrary multivariate function, $g : \mathbb{R}^{d} \rightarrow \mathbb{R}$, we follow Ref.~\cite{DIRT2022} and discretise $g$ using a tensor-product grid, then compute a low-rank approximation to the resulting tensor in tensor train format using the cross approximation algorithm developed in Ref. \cite{oseledets2010tt}. We then use a set of basis functions to interpolate between the values of the tensor at each grid point. The resulting FTT approximation takes the form
\begin{equation}
    g(\mathbf{x}) \approx \mathsf{G}_{1}(x_{1})\mathsf{G}_{2}(x_{2}) \cdots \mathsf{G}_{d}(x_{d}), \label{eq:ftt}
\end{equation}
where each $\mathsf{G}_{k} : \mathbb{R} \rightarrow \mathbb{R}^{r_{k-1} \times r_{k}}$ denotes a matrix-valued function, and the scalars $\{r_{k}\}_{k=1}^{d}$ are referred to as the tensor train ranks (note that $r_{0}=r_{d}=1$). Such a representation can be constructed efficiently; under the assumption that the maximum rank, $r = \max_{k} r_{k}$, and the maximum number of gridpoints in each dimension, $n$, remain bounded as the dimension of the function increases, it follows that forming the approximation of Eq.~\eqref{eq:ftt} requires $\mathcal{O}(dnr^{2})$ evaluations of the target function.

To construct an FTT approximation to an arbitrary probability density function, $\pi : \mathbb{R}^{d} \rightarrow \mathbb{R}$, we compute an FTT approximation, $\hat{g}$, to the \emph{square root} of the ratio between $\pi$ and a positive weighting function $\lambda$ (which is specific to the choice of approximation basis; see, e.g., \cite{cui2023selfreinforced}); that is, $\hat{g} \approx \sqrt{\pi/\lambda}$. The resulting approximation to $\pi$ takes the form
\begin{equation}
    \pi(\boldsymbol{\theta}) \approx \hat{\pi}(\boldsymbol{\theta}) = \frac{1}{\hat{z}}(\hat{g}(\boldsymbol{\theta})^{2} + \tau)\lambda(\boldsymbol{\theta}), \quad \hat{z} := \int_{\boldsymbol{\Theta}} (\hat{g}(\boldsymbol{\theta})^{2} + \tau) \lambda(\boldsymbol{\theta}) \, \mathrm{d}\boldsymbol{\theta}. \label{eq:sirt}
\end{equation}
In Eq.~\eqref{eq:sirt}, $\tau > 0$ denotes a small constant, used to ensure that $\hat{\pi}$ is a valid importance sampling density \cite{DIRTRareEvent2024, DIRT2022}, which can be chosen based on the $L^{2}$ error of $\hat{g}$. By computing an FTT approximation to the square root of the ratio between $\pi$ and $\lambda$, we avoid that Eq.~\eqref{eq:sirt} takes on negative values \cite{DIRT2022} (note that Eq.~\eqref{eq:sirt} is positive everywhere by construction).

To sample from the approximation $\hat{\pi}$, we can construct the Rosenblatt transport (also referred to as the Knothe-Rosenblatt rearrangement) \cite{knothe1957contributions, rosenblatt1952remarks}, which provides a deterministic mapping between a random variable distributed according to $\hat{\pi}$, and the $d$-dimensional uniform random variable, with density denoted by $\mu$. To define the Rosenblatt transport, we start by defining the marginal cumulative distribution function (CDF) of the first component of $\boldsymbol{\theta}$, denoted by $\theta_{1}$, which takes the form
\begin{equation}
    F_{1}(\theta_{1}) := \int_{-\infty}^{\theta_{1}}\hat{\pi}_{1}(\theta_{1}') \, \mathrm{d}\theta_{1}',
\end{equation}
where $\hat{\pi}_{1}$ denotes the approximation to the marginal density of $\theta_{1}$. Additionally, we define the sequence of conditional CDFs
\begin{equation}
    F_{k|<k}(\theta_{k} | \boldsymbol{\theta}) := \int_{-\infty}^{\theta_{k}} \hat{\pi}_{k|<k}(\theta_{k}' | \boldsymbol{\theta}_{<k}) \, \mathrm{d}\theta_{k}', \quad k = 2, 3, \dots, d,
\end{equation}
where $\boldsymbol{\theta}_{<k}$ is shorthand for $[\theta_{1}, \theta_{2}, \dots, \theta_{k-1}]^{\top}$, and $\hat{\pi}_{k|<k}$ denotes the approximation to the density of $\theta_{k}|\boldsymbol{\theta}_{<k}$. Finally, the Rosenblatt transport is defined as 
\begin{equation}
    \mathcal{F}(\boldsymbol{\theta}) := \begin{bmatrix}
        F_{1}(\theta_{1}) \\
        F_{2|<2}(\theta_{2}|\boldsymbol{\theta}_{<2}) \\
        \vdots \\
        F_{d|<d}(\theta_{d}|\boldsymbol{\theta}_{<d})
    \end{bmatrix}, \label{eq:rosenblatt}
\end{equation}
and satisfies the relation $\mathcal{F}_{\sharp}\hat{\pi} = \mu$. For many choices of basis functions and weighting function $\lambda$, the marginal and conditional CDFs defining the Rosenblatt transport can be evaluated in closed form \cite{cui2023selfreinforced}. Additionally, because Eq.~\eqref{eq:sirt} is strictly positive, each component of the corresponding Rosenblatt transport is strictly monotone and thus invertible. The inverse Rosenblatt transport, $\mathcal{F}^{-1}$, can be evaluated by solving a sequence of one-dimensional root-finding problems using a standard technique such as Newton's method.

It is possible to extend the Rosenblatt transport to map from the FTT approximation \eqref{eq:sirt} to an arbitrary product-form reference density, $\rho$. First, note that the coupling
\begin{equation}
    \mathcal{R}(\boldsymbol{u}) := \left[R_{1}(u_{1}), R_{2}(u_{2}), \dots, R_{d}(u_{d})\right]^{\top}, \quad R_{k}(u_{k}) := \int_{-\infty}^{u_{k}} \rho_{k}(u_{k}') \, \mathrm{d}u_{k}',
\end{equation}
satisfies $\mathcal{R}_{\sharp} \rho = \mu$. The associated (general) inverse Rosenblatt transport, denoted by $\mathcal{Q} := \mathcal{F}^{-1} \circ \mathcal{R}$, satisfies $\mathcal{Q}_{\sharp}\rho = \hat{\pi}$; this implies that we can sample from $\hat{\pi}$ by applying the mapping $\mathcal{Q}$ to samples drawn from the reference density $\rho$. Throughout this work, we use a standard $d$-dimensional Gaussian density as the reference density.

In situations where the target density is concentrated within a small region of the parameter space or exhibits a complex correlation structure, it can be challenging to construct an FTT approximation of the form of Eq.~\eqref{eq:sirt} directly. To address this challenge, \cite{DIRT2022} construct a composition of mappings which gradually adapt towards the target density; the resulting \emph{deep} inverse Rosenblatt transport is denoted by $\mathcal{T} = \mathcal{Q}^{(1)} \circ \mathcal{Q}^{(2)} \circ \cdots \circ \mathcal{Q}^{(K)}$.

The construction of the DIRT is guided by a sequence of bridging densities, denoted by $\phi^{(1)}, \phi^{(2)}, \dots, \phi^{(K)} := \pi$, which interpolate between a density that is simple to construct an FTT approximation to, and the target density. The $k$th mapping is constructed to approximate the pullback of the $k$th bridging density under the incremental mapping $\mathcal{T}^{(k-1)} := \mathcal{Q}^{(1)} \circ \dots \circ \mathcal{Q}^{(k-1)}$, which is given by 
\begin{equation}\label{eq:aratio}
    \begin{aligned}
            (\mathcal{T}^{(k-1)})^{\sharp} \phi^{(k)}(\boldsymbol{u}) &= \phi^{(k)}(\mathcal{T}^{(k-1)}(\boldsymbol{u}))|\nabla_{\boldsymbol{u}}\mathcal{T}^{(k-1)}(\boldsymbol{u})| \\
    &= \frac{\phi^{(k)}(\mathcal{T}^{(k-1)}(\boldsymbol{u}))}{\hat{\phi}^{(k-1)}(\mathcal{T}^{(k-1)}(\boldsymbol{u}))}\rho(\boldsymbol{u}) \\
    &\approx \frac{\phi^{(k)}(\mathcal{T}^{(k-1)}(\boldsymbol{u}))}{\phi^{(k-1)}(\mathcal{T}^{(k-1)}(\boldsymbol{u}))}\rho(\boldsymbol{u}), 
    \end{aligned}
\end{equation}
where $\hat{\phi}^{(k-1)}$ denotes the FTT-based approximation to bridging density $k-1$, and the second line uses the identity
\begin{equation}
    (\mathcal{T}^{(k-1)})^{\sharp}\hat{\phi}^{(k-1)}(\boldsymbol{u}) = \hat{\phi}^{(k-1)}(\mathcal{T}^{(k-1)}(\boldsymbol{u}))|\nabla_{\boldsymbol{u}}\mathcal{T}^{(k-1)}(\boldsymbol{u})| = \rho(\boldsymbol{u}).
\end{equation}
If successive bridging densities are chosen such that they are similar to one another, it should be simpler to construct an FTT approximation to each of the (approximate) pullback densities of the form of Eq.~\eqref{eq:aratio} than to the target density directly.

When constructing a DIRT mapping to approximate the optimal biasing density $q^{y*}$ (recall that this coincides with the posterior density, $\pi^{y}$) in the \emph{a posteriori} setting, we can specify the bridging densities using a tempering approach (see, e.g., \cite{del2006sequential}). In this setting, bridging density $k$ takes the form 
\begin{equation}
    \phi^{(k)}(\boldsymbol{\theta}) \propto \pi_{0}(\boldsymbol{\theta})^{1-\beta_{k}}\pi^{y}(\boldsymbol{\theta})^{\beta_{k}},
\end{equation}
where $\pi_{0}$ is a density that is simple to construct an FTT approximation to (typically the prior), and the coefficients $\{\beta_{k}\}_{k=1}^{K}$ are defined such that $0 \leq \beta_{1} < \beta_{2} < \cdots < \beta_{K} = 1$. When approximating the optimal biasing densities $p^{*}$ (in the \emph{a priori} setting) or $p^{y*}$ (in the \emph{a posteriori} setting), however, we take a slightly different approach. Note that the presence of the indicator function for the failure region will introduce discontinuities into both $p^{*}$ and $p^{y*}$, which can result in the tensor train ranks required to approximate both of these densities becoming large \cite{DIRTRareEvent2024}. To address this, we replace the indicator function, $\mathbb{I}_{\mathcal{F}}(\boldsymbol{\theta})$, with a smooth surrogate, $f(\boldsymbol{\theta}; \gamma)$, parametrized by a scalar, $\gamma$, such that $\lim_{\gamma\rightarrow\infty}f(\boldsymbol{\theta}; \gamma) = \mathbb{I}_{\mathcal{F}}(\boldsymbol{\theta})$. In the present work, we follow  Ref.~\cite{DIRTRareEvent2024} and use a sigmoid function which takes the form
\begin{equation}
    f(\boldsymbol{\theta}; \gamma) = (1 + \exp[\gamma(z-F(\boldsymbol{\theta}))])^{-1}.
\end{equation}
When approximating $p^{*}$, to construct a sequence of bridging densities we vary the parameter $\gamma$ between a sufficiently small value and the eventual target value, $\gamma^{*}$. In this setup, bridging density $k$ is given as
\begin{equation}
    \phi^{(k)}(\boldsymbol{\theta}) \propto f(\boldsymbol{\theta}; \gamma^{(k)}) \pi_{\mathrm{pr}}(\boldsymbol{\theta}),
\end{equation}
where $0 < \gamma^{(1)} < \gamma^{(2)} < \cdots < \gamma^{(K)} = \gamma^{*}$. When approximating $p^{y*}$, we may wish to apply additional tempering to the posterior; in this case, bridging density $k$ takes the form
\begin{equation}
    \phi^{(k)}(\boldsymbol{\theta}) \propto f(\boldsymbol{\theta}; \gamma^{(k)}) \pi_{0}(\boldsymbol{\theta})^{1-\beta_{k}}\pi^{y}(\boldsymbol{\theta})^{\beta_{k}},
\end{equation}
where the coefficients $\{\beta_{k}\}_{k=1}^{K}$ are defined as before.

\begin{figure}
    \centering
    \includegraphics[width=0.95\linewidth]{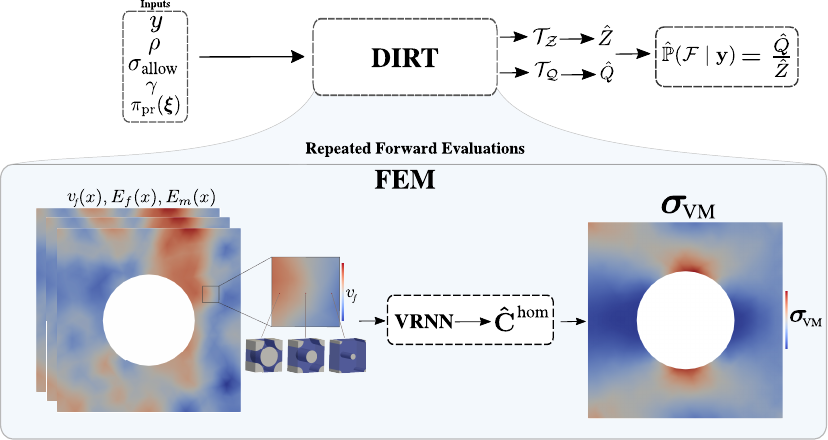}
    \caption{Multiscale reliability analysis framework with DIRT and VRNN.}
    \label{fig:framework_dirt_vrnn}
\end{figure}

The complete importance sampling framework with DIRT and VRNN is summarized in Figure~\ref{fig:framework_dirt_vrnn}.

\section{Numerical Example}
\label{sec:example}
\paragraph{Problem Setup and Objectives}
The goal of this section is to assess whether the proposed multiscale framework can accurately and with low variance estimate posterior failure probabilities as the input dimension increases. The plate-with-hole benchmark problem is selected because failure is governed by stress concentrations around the hole and is highly sensitive to material heterogeneity. At each macroscale integration point, the uncertain microscale parameters $(v_f(x), E_f(x), E_m(x))$ are mapped by the VRNN surrogate to a homogenized constitutive tensor $\hat{\mathbf{C}}^{\text{hom}}(x)$, which is then used in the finite element solution of the macroscale forward problem. As a verification of the DIRT-based posterior sampler, we also study the quality of the DIRT approximation to the posterior of the uncertain microscale parameters based on noisy measurement data.  

\subsection{Model Description}
\begin{figure}[h!]
    \centering
    \begin{subfigure}[b]{0.52\linewidth}
        \centering
        \includegraphics[width=\linewidth]{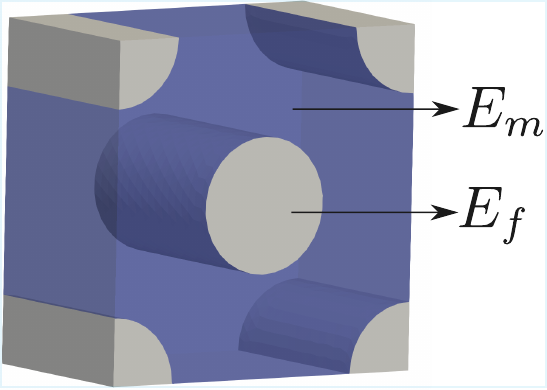}
        \caption{Representative Volume Element with unidirectional fibers. Microscale parameters ($v_f, E_f,E_m$) define the phase properties.}
        \label{fig:RVE_domain}        
    \end{subfigure}
    \hfill
    \begin{subfigure}[b]{0.40\linewidth}
        \centering
        \includegraphics[width=\linewidth]{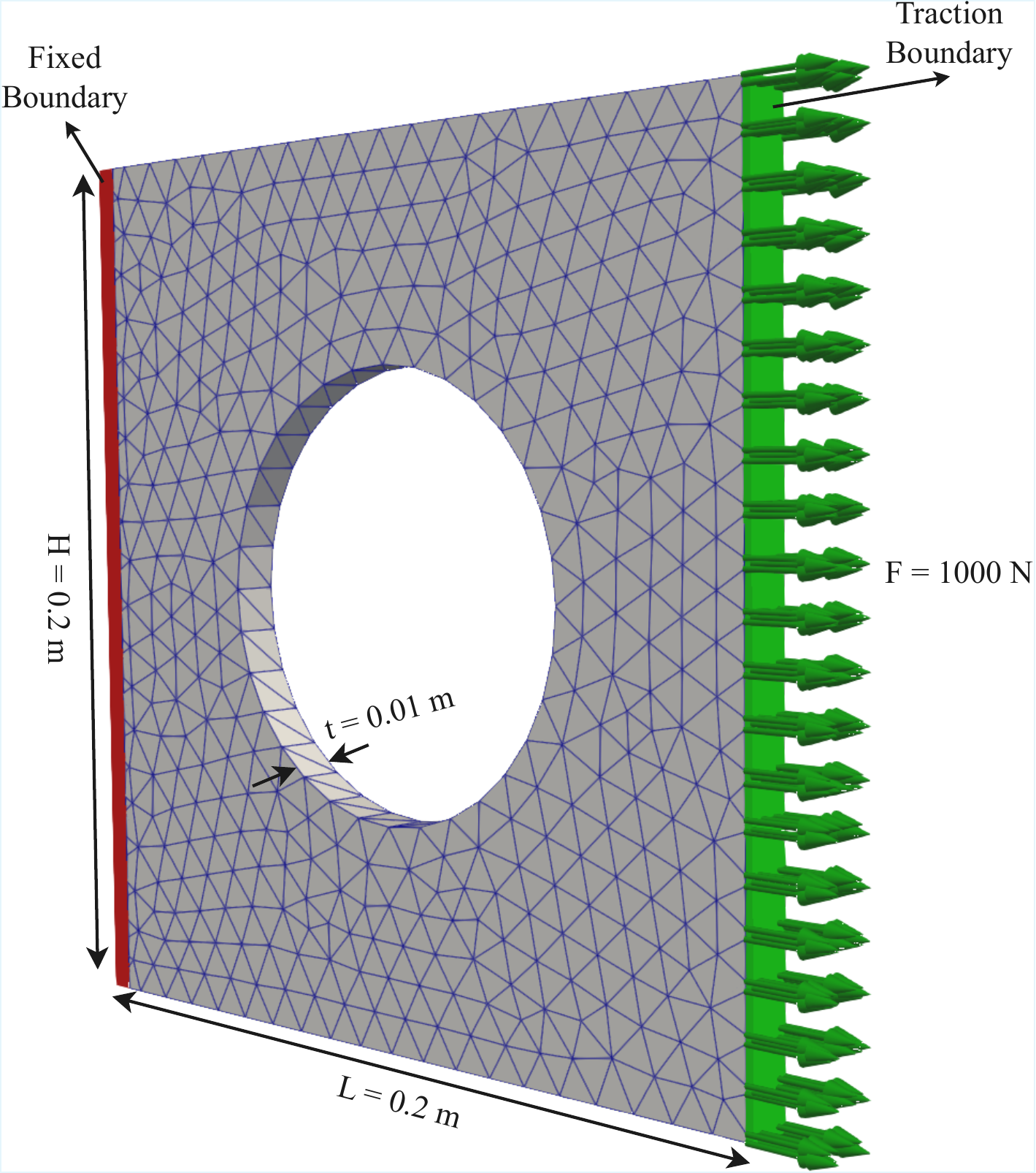}
        \caption{Macroscale plate-with-hole finite-element mesh  and boundary conditions}
        \label{fig:macro_domain}
    \end{subfigure}
\label{fig:domains}
\caption{Microscale and macroscale geometries used in the numerical experiment.}
\end{figure}

\begin{figure}[h!]
    \centering
    \includegraphics[width=\linewidth]{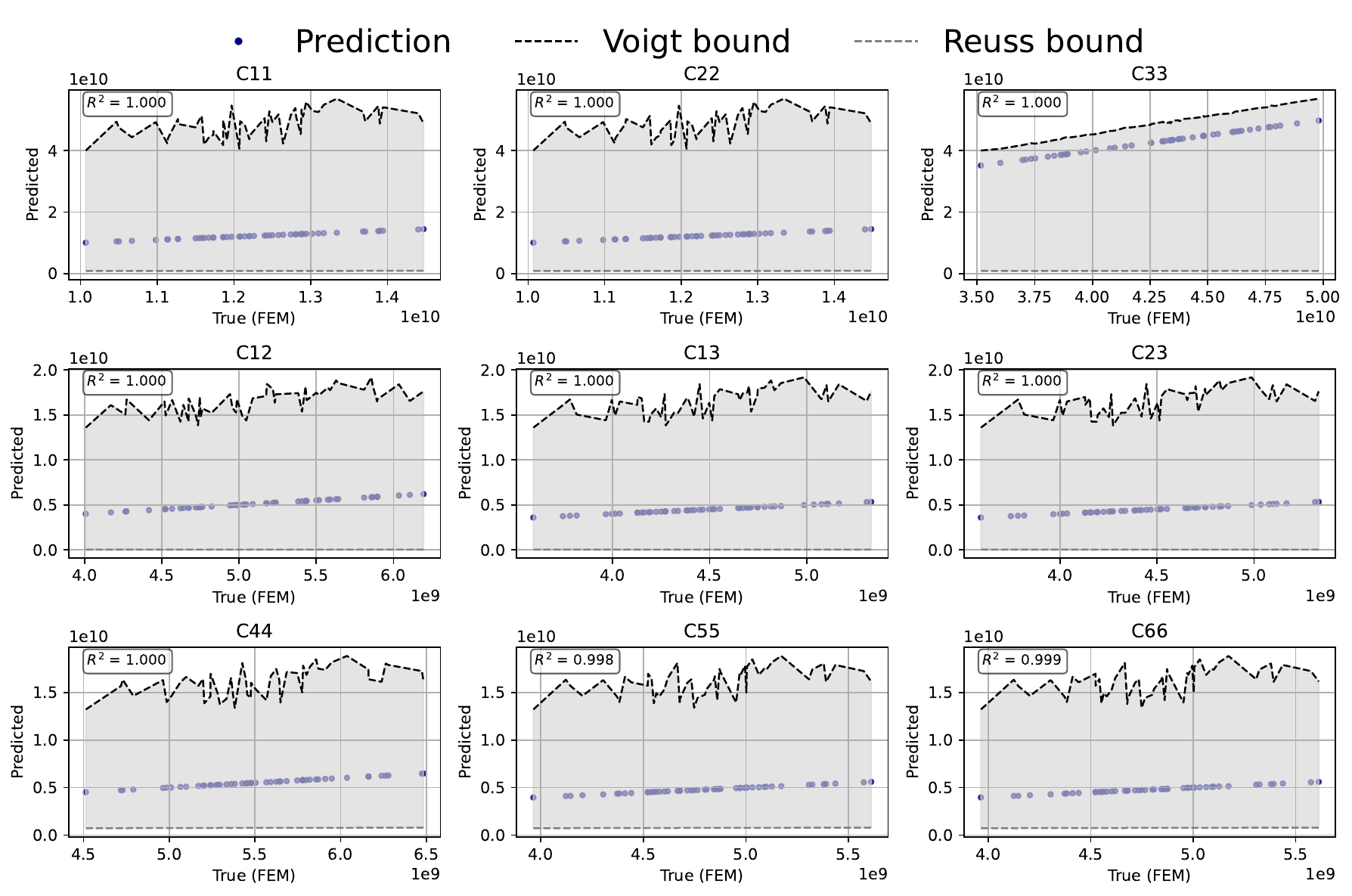}
    \caption{Comparison of homogenized stiffness tensor components predicted by VRNN with reference finite element (FEM) results. 
    Each subplot corresponds to one independent entry of the homogenized stiffness tensor $\mathbf{C}^{\text{hom}}$ in Voigt notation. 
    Blue markers denote VRNN predictions, while the dashed black and gray lines indicate the theoretical Voigt and Reuss bounds, respectively.}
    \label{fig:VRNN_bounds}
\end{figure}

\subsubsection{Microscale RVE Model}
The microstructure is represented with an RVE containing unidirectional fibers as shown in Fig.~\ref{fig:RVE_domain}. The RVE is parameterized by the fiber volume fraction $v_f$, the Young's modulus of the fibers $E_f$, and the Young's modulus of the matrix $E_m$. The data set comprises $N=250$ RVE realizations, where the parameters are sampled uniformly in the ranges $v_f \in [0.4,\,0.7]$, $E_f \in [50,\,80]$ GPa, and $E_m \in [2,\,5]$ GPa. The homogenized stiffness tensors are obtained through FE$^2$ simulations using the \texttt{python}-based package \texttt{sfepy} \cite{sfepy}. Each RVE simulation solves the RVE elasticity problem under periodic boundary conditions on the displacement fluctuations corresponding to an imposed macroscopic strain. The obtained $\mathbf{C}^{\text{hom}}\in\mathbb{R}^{6\times6}$ (Voigt notation) is used as the ground truth and the dataset is split randomly in an 80–20 ratio for training and validation.
\begin{figure}[h!]
    \centering
    \includegraphics[width=0.5\linewidth]{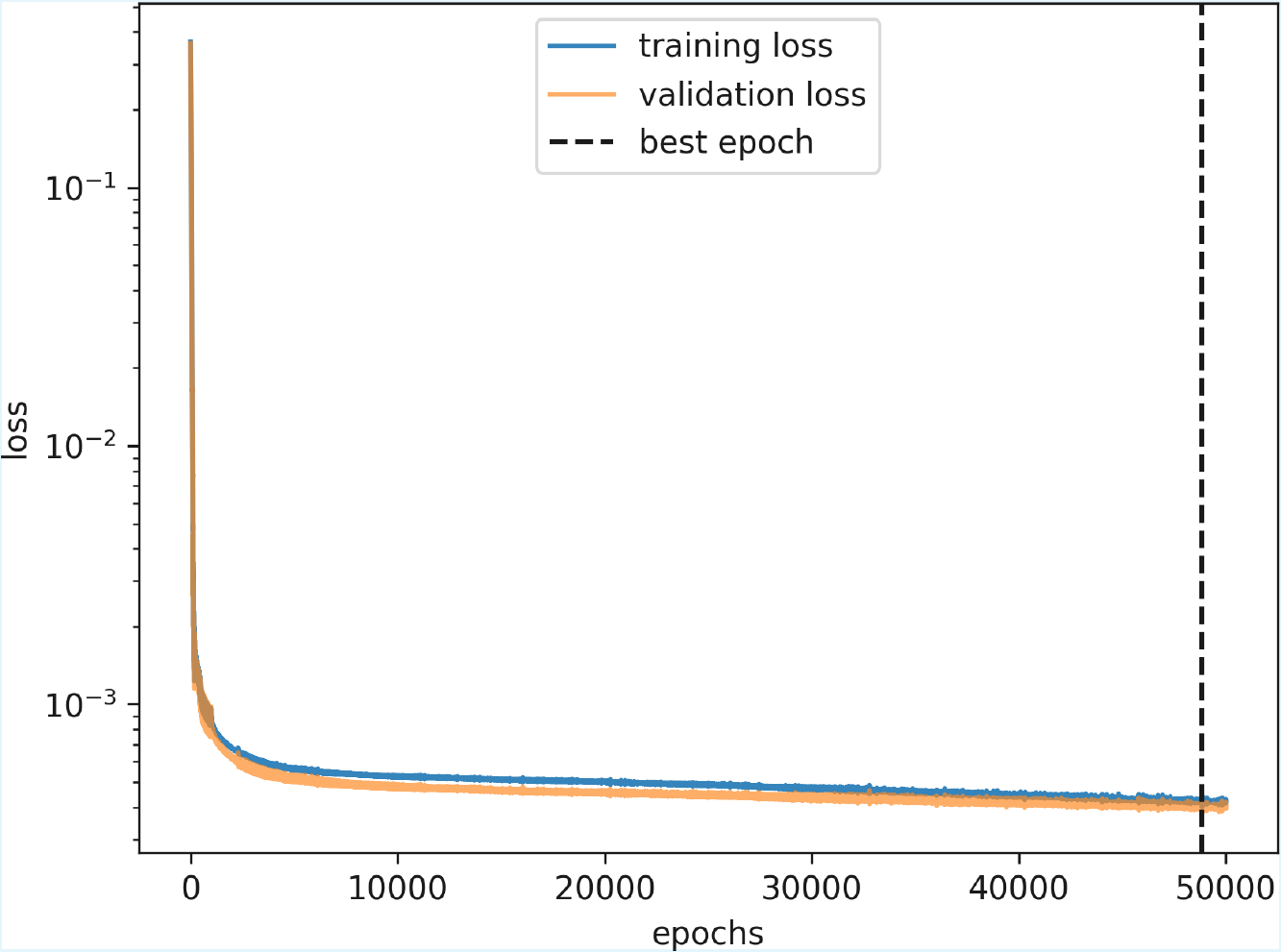}
    \caption{Training and validation loss evolution for VRNN.}
    \label{fig:VRNN_loss}
\end{figure}
In the following, we employ the same VRNN architecture (number of layers, etc.) as proposed in Ref.~\cite{keshav2025spectral}. We remark that the final output layer employs a \texttt{sigmoid} activation to restrict $\xi_\lambda \in (0,1)$. The \texttt{AdamW} optimizer is used for training with an initial learning rate of $10^{-4}$ and a weight decay coefficient of $10^{-4}$. The network is trained for 50,000 epochs and the training history is shown in Figure~\ref{fig:VRNN_loss}. VRNN ensures that each predicted homogenized stiffness tensor $\hat{\mathbf{C}}^{\text{hom}}$ is strictly bounded within the Voigt–Reuss limits (see Figure~\ref{fig:VRNN_bounds}). The figure illustrates the comparison between VRNN predictions and FEM homogenization results for all independent entries of the stiffness tensor. The predicted values consistently remain within the theoretical Voigt-Reuss bounds. All predicted components also exhibit strong correlation with the true (obtained through FEM homogenization) components as indicated by the high coefficients of correlation ($R^2\approx1)$.

\subsubsection{Macroscale Model}

\begin{figure}[h!]
    \centering
    \begin{subfigure}[b]{0.32\linewidth}
        \centering
        \includegraphics[width=\linewidth]{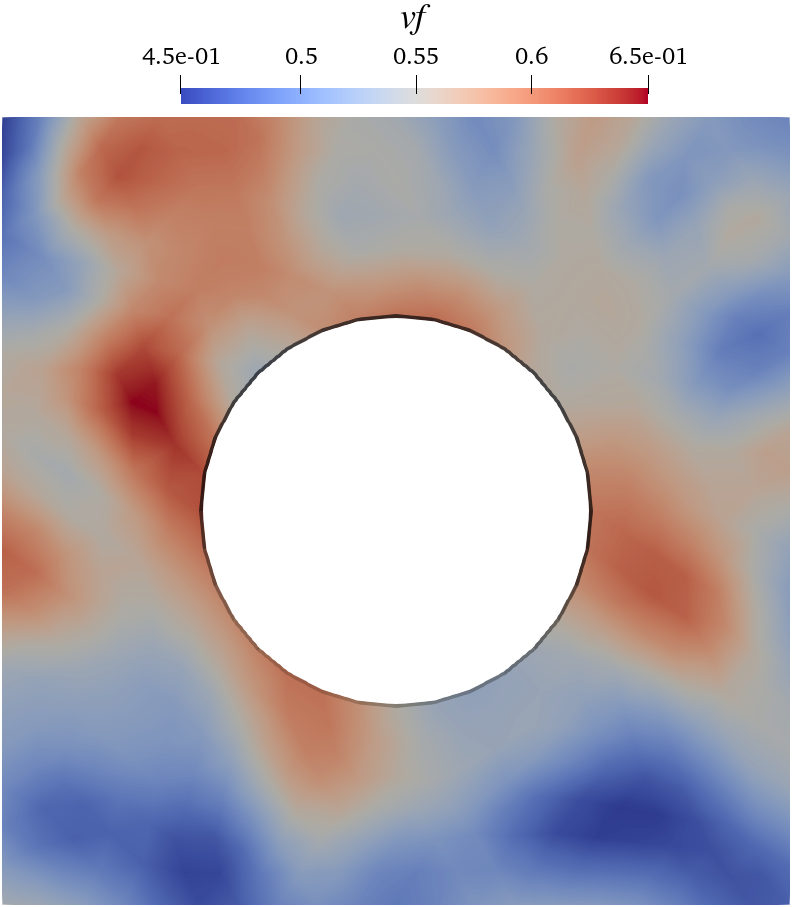}
        \caption{Volume Fraction}
        \label{fig:a}
    \end{subfigure}
    \hfill
    \begin{subfigure}[b]{0.32\linewidth}
        \centering
        \includegraphics[width=\linewidth]{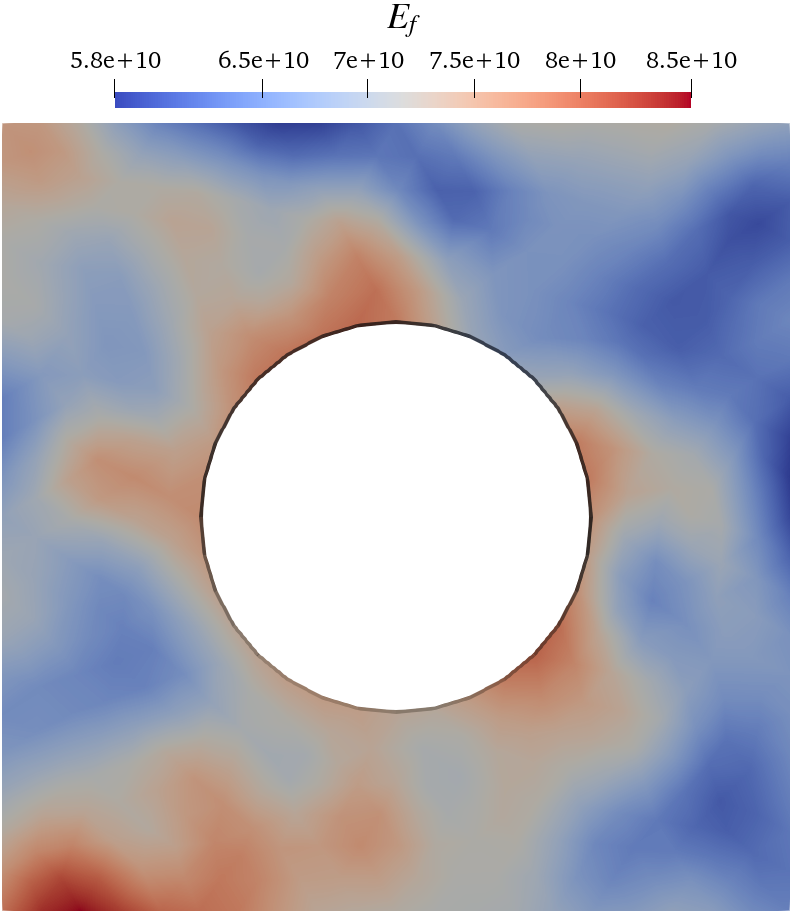}
        \caption{Fiber Young's Modulus}
        \label{fig:b}
    \end{subfigure}
    \hfill
    \begin{subfigure}[b]{0.32\linewidth}
        \centering
        \includegraphics[width=\linewidth]{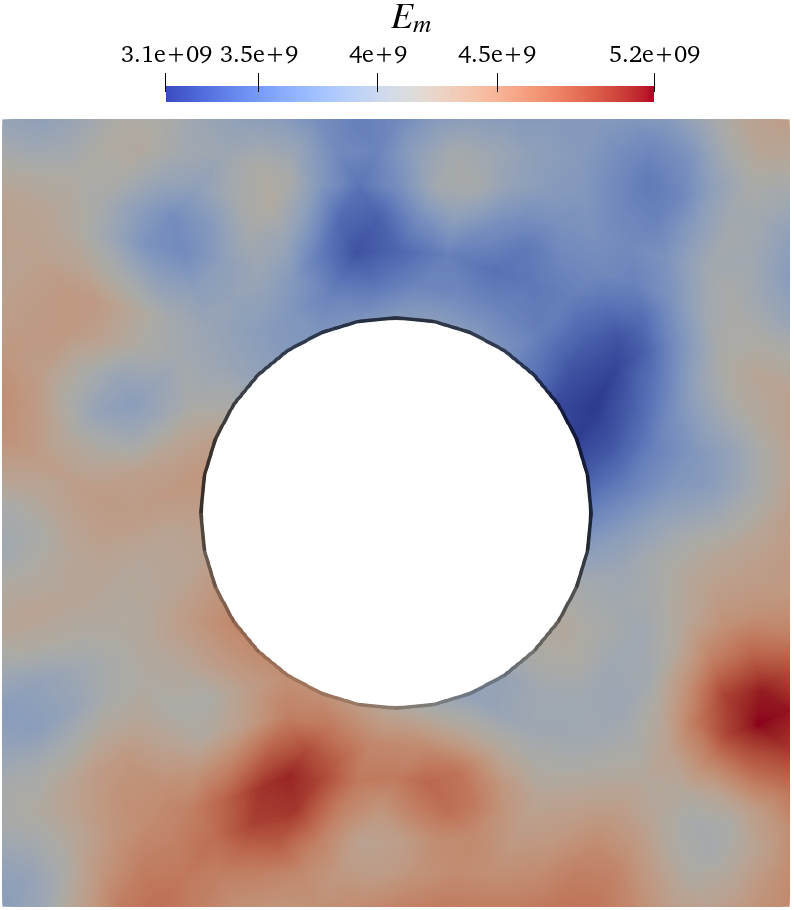}
        \caption{Matrix Young's Modulus}
        \label{fig:c}
    \end{subfigure}
    \caption{Realizations of the random fields}
    \label{fig:realizations}
\end{figure}

For the macroscale problem, we consider a thin plate with a central hole made of a fiber-reinforced composite material (Fig.~\ref{fig:macro_domain}). The left boundary is fixed, while a uniform tensile force of $F = 1000$~N is applied on the right edge. The finite element model of the plate consists of 2,703 tetrahedral elements. For each realization of the random fields, the equilibrium equations are solved using a linear solver.
The RVE parameters $v_f$, $E_f$, and $E_m$ are treated as spatially random fields over the plate domain, representing material uncertainty due to microscale variability. 
The prior random fields are modeled using a log-normal distribution with the statistics summarized in Table~\ref{tab:random_field_stats}.
One realization of each field drawn from the prior is shown in Fig.~\ref{fig:realizations}.

\begin{table}[h!]
    \centering
    \caption{Statistical properties of the prior random fields for the three RVE parameters.}
    \label{tab:random_field_stats}
    \begin{tabular}{lcccc}
        \toprule
        \textbf{Parameter} & \textbf{Mean ($\mu$)} & \textbf{Std Dev} & \textbf{Correlation length ($\ell$)} & \textbf{Distribution} \\
        \midrule
        $v_f$ [-] & 0.52 & 5\% & 0.05 & Log Normal \\
        $E_f$ [GPa] & 72.45 & 5\% & 0.05 & Log Normal \\
        $E_m$ [GPa] & 3.43 & 5\% & 0.05 & Log Normal \\
        \bottomrule
    \end{tabular}
\end{table}

\begin{figure}
    \centering
    \includegraphics[width=0.5\linewidth]{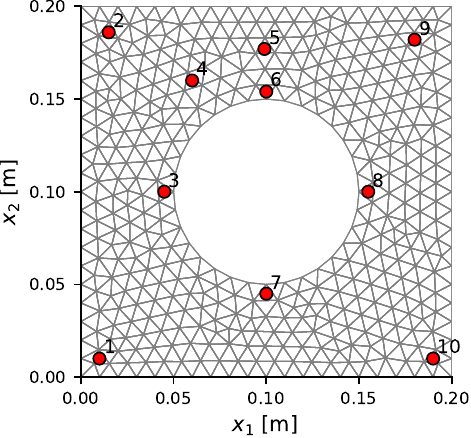}
    \caption{Finite element mesh of the plate with hole showing the ten locations (red markers) used for strain measurements in the Bayesian inversion problem.}
    \label{fig:observed_data}
\end{figure}

To establish confidence in the homogenization surrogate, validation of the VRNN-driven macroscale response against classical FE$^2$ homogenization is performed in Appendix~\ref{appendix:vrnn_validation}. The results show that the VRNN-based homogenized response demonstrates close agreement in the resulting macroscale displacement and stress fields. 

\paragraph{DIRT hyperparameters.}
The construction of the DIRT transport map is controlled by four main hyperparameters: (i) the basis size $n$ used to parameterize each one-dimensional conditional, (ii) the number of transport layers, $L$, (iii) the maximum tensor train rank $r$ used in the low-rank representation, and (iv) the smoothing parameter $\gamma^{*}$ used in the smooth surrogate of the failure function. Following practical guidance and hyperparameters studied in Ref.~\cite{DIRTRareEvent2024}, we fix $n=30$ and $L=12$ across all dimensionalities. We tuned and chose $\gamma^{*}=50$ to ensure stable training and well-behaved importance weights in the failure function transport map. The remaining hyperparameter, the maximum TT rank $r$, directly controls the trade-off between approximation accuracy and computational cost and is therefore adjusted with dimension and target function complexity. Because each function evaluation requires a full macroscale FE solve, exhaustive hyperparameter search is computationally infeasible. Instead, we determine $r$ via a budgeted calibration procedure: starting from a conservative rank $r = 2$, we increase $r$ until improvements in transport quality and estimator stability (e.g., higher effective sample size / lower weight variance and stable $\Pfpostest$ across repeated runs) saturate within the available computational budget. The final ranks for each dimension are reported along with the failure probability estimates in Section~\ref{sec:pf}.

\subsection{Results}

\subsubsection{Bayesian Inference and Field Reconstruction}

\begin{figure}
    \centering
    \includegraphics[width=0.85\linewidth]{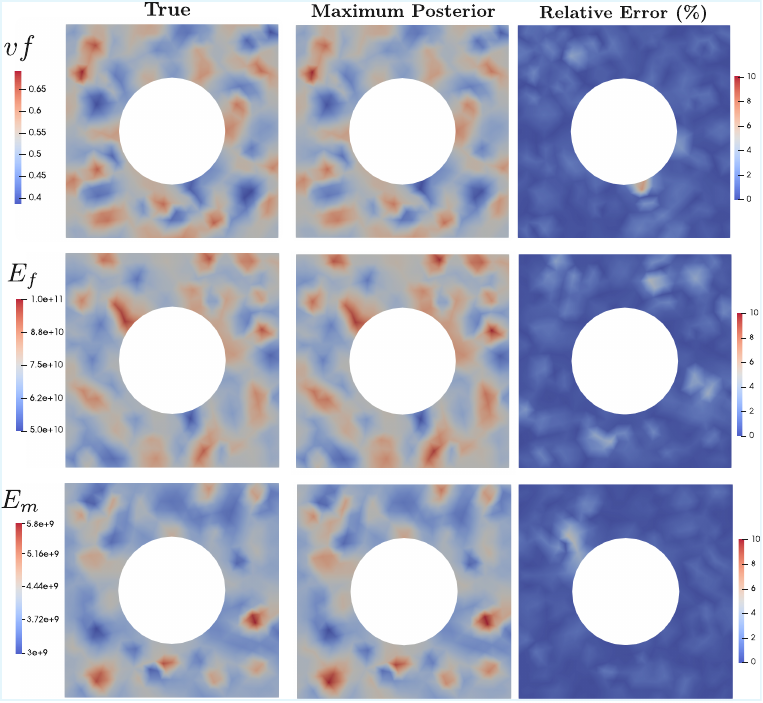}
    \caption{Comparison of the true, reconstructed, and relative error fields for \(v_f\) (top row), \(E_f\) (middle row), and \(E_m\) (bottom row) for $M = 50$.}
    \label{fig:150D_post_compare}
\end{figure}

To estimate the "true" microscale fields underlying the macroscale response, a Bayesian inverse problem is formulated based on noisy strain measurements obtained at $N_{\text{obs}} = 10$ locations on the plate (Fig.~\ref{fig:observed_data}). The measurement noise is assumed to be additive, independent, and Gaussian, with a standard deviation of $\sigma_{\text{noise}} = 10^{-5}$. Let $\mathbf{y}_{\text{obs}} \in \mathbb{R}^{20}$ denote the observed strain data, which represent the in-plane normal strain components ($\epsilon_{xx}$ and $\epsilon_{yy}$) at the selected measurement points.
The random-field parameterization of $\{v_f, E_f, E_m\}$ and the definition of the KL coefficients are given in Section~\ref{sec:reliability}. $\boldsymbol{\xi}\in\mathbb{R}^d$ denotes the concatenated KL coefficients for the random fields ($d = 3\times M$ when using $M$ KL terms per field). The goal of the inference is to recover the heterogeneous microscale quantities $\{v_f(\mathbf{x}), E_f(\mathbf{x}), E_m(\mathbf{x})\}$. 
The likelihood function is defined by assuming Gaussian observational noise:
\begin{equation}
    \mathcal{L}(\mathbf{y}_{\text{obs}} | \boldsymbol{\xi})
    \propto
    \exp\left( 
        -\frac{1}{2\sigma_{\text{noise}}^2}
        \| \mathbf{y}_{\text{obs}} - \mathbf{y}_{\text{model}}(\boldsymbol{\xi}) \|_2^2
    \right),
\end{equation}

where $\mathbf{y}_{\mathrm{model}}(\boldsymbol{\xi})$ denotes the predicted strain measurements obtained from the forward finite element model based on the VRNN-derived stiffness fields. In particular, for a given $\boldsymbol{\xi}$, the fields $\{v_f(\mathbf{x}), E_f(\mathbf{x}), E_m(\mathbf{x})\}$ are reconstructed and mapped by the VRNN to the homogenized stiffness field $\mathbf{C}^{\mathrm{hom}}(\mathbf{x})$. The resulting macroscale FE problem is then solved, and the strain components at the $N_{\mathrm{obs}}$ sensor locations are extracted to form $\mathbf{y}_{\mathrm{model}}(\boldsymbol{\xi})$. A standard normal prior is assigned to the KL coefficients,

\begin{equation}
p(\boldsymbol{\xi}) = \mathcal{N}(\mathbf{0}, \mathbf{I}_d),    
\end{equation}
and the posterior distribution follows from Bayes’ theorem:
\begin{equation}
p(\boldsymbol{\xi} | \mathbf{y}_{\text{obs}}) \propto 
\mathcal{L}(\mathbf{y}_{\text{obs}} | \boldsymbol{\xi}) \, p(\boldsymbol{\xi}).    
\end{equation}
The posterior mean of the KL coefficients, $\mathbb{E}[\boldsymbol{\xi} | \mathbf{y}_{\text{obs}}]$, is used to reconstruct the approximated posterior mean fields of the RVE parameters
\begin{equation}
    \hat{\boldsymbol{\theta}}(\mathbf{x}) = 
    \bar{\boldsymbol{\theta}}(\mathbf{x}) + \sum_{i=1}^{d} \sqrt{\lambda_i}\, \boldsymbol{\phi}_i(\mathbf{x})\, 
    \mathbb{E}[\xi_i | \mathbf{y}_{\text{obs}}] \ .
\end{equation}
Figure~\ref{fig:150D_post_compare} shows the comparison between the "true" fields and the approximated mean posterior fields for each microstructure descriptor, along with the relative errors. It can be observed that the relative errors are within $10\%$ for all three fields.

\subsubsection{Reliability Analysis}
\label{sec:pf}

For the reliability analysis, the maximum von Mises stress in the plate, denoted by $\max\sigma_{\mathrm{VM}}(\boldsymbol{\theta})$, is chosen to be the quantity of interest. The performance function is thus defined as
\begin{equation}
g(\boldsymbol{\theta}) = \sigma_{\mathrm{allow}} - \max\sigma_{\mathrm{VM}}(\boldsymbol{\theta}) \ ,    
\end{equation}
with $\sigma_{\mathrm{allow}} = 0.96 \, \text{MPa}$ being the prescribed failure threshold. 
\begin{figure}[h!]
    \centering
    \includegraphics[width=0.75\linewidth]{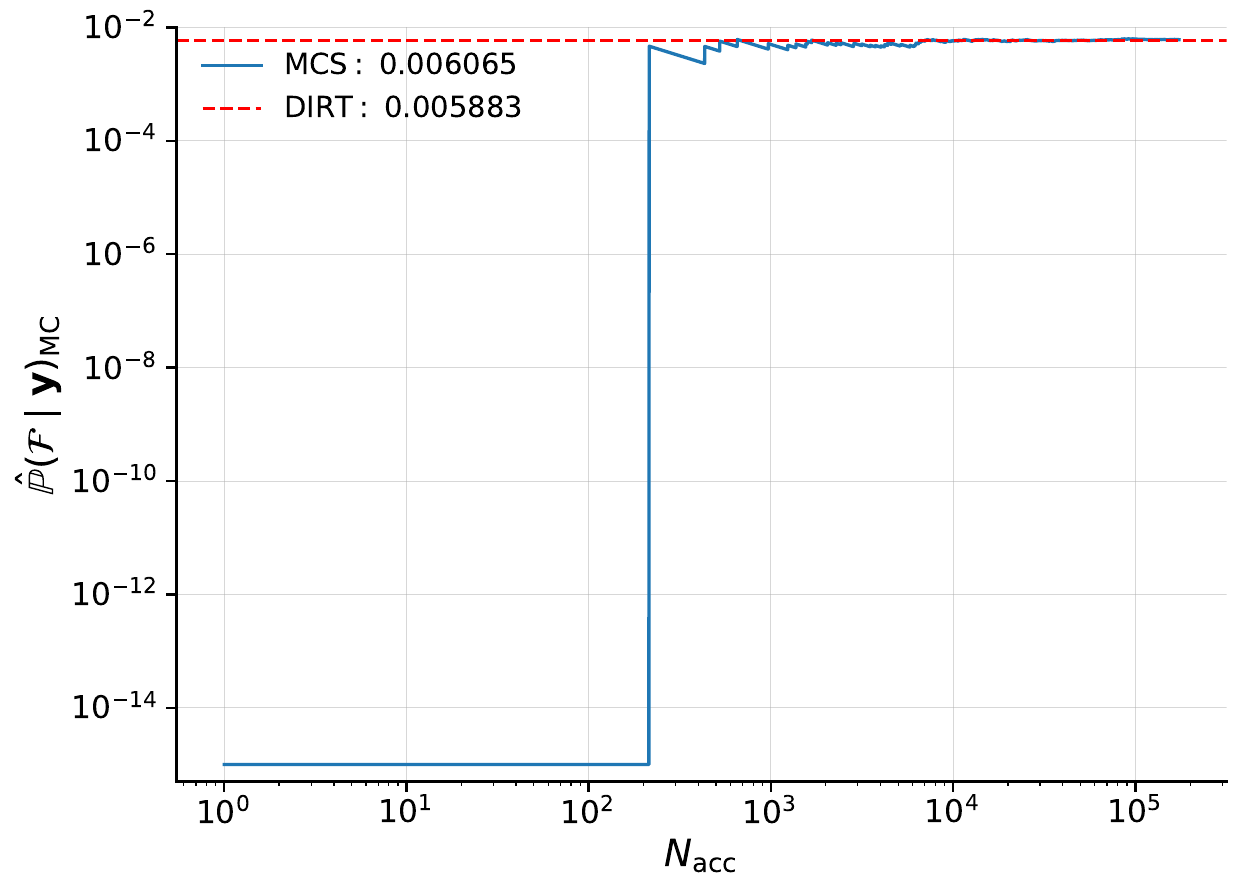}
    \caption{Running estimate of $\Pfpostest_{\text{MC}}$ with increasing $N_{\text{acc}}$. For visualization, values of $\Pfpostest_{\text{MC}} = 0$ are displayed as $10^{-15}$ in the plot.} 
    \label{fig:pf_mcs}
\end{figure}

\paragraph{12 Dimensional Monte Carlo reference} Before scaling the DIRT algorithm to higher dimensions, we computed a reference posterior failure probability $\Pfpostest_{\text{MC}}$ for $d = 12$.  
For obtaining samples from the posterior, we generated $N_{\text{prop}} = 9.6\times10^8$ samples from the prior with 20 independent seeds ($4.8\times10^7$ samples per seed) on the Texas Advanced Computing Center (TACC), and $N_\text{acc} = 167,000$ samples were accepted in the posterior, corresponding to an acceptance rate of $0.0174 \%$. $\Pfpostest_{\text{MC}}$ was then computed as 
\begin{equation}
    \Pfpostest_{\text{MC}} = \frac{1}{N_{\text{acc}}}\sum_{i=1}^{N_{\text{acc}}}\mathbb{I}\{g(\boldsymbol{\xi}^{(i)})\leq0\} \ .
\end{equation}
The resulting $\Pfpostest_{\text{MC}}$ is $\approx6.06\times10^{-3}$. The convergence of $\Pfpostest_{\text{MC}}$ with increasing posterior samples is shown in Figure~\ref{fig:pf_mcs}. The close agreement between the two estimators demonstrates that DIRT is able to accurately estimate posterior failure probabilities. As shown in Table~\ref{tab:pf_post_summary}, for the 12D case, DIRT required only 311,000 forward FEM evaluations to obtain roughly the same value of $\Pfpostest$ as MCS.
\begin{table}[h!]
    \centering
    \caption{Posterior failure probability estimates across dimensions with 3 independent runs. Here, $N_{\text{evals}}$ denotes the total number of FEM solutions used to build the DIRT transport maps.}
    \label{tab:pf_post_summary}
    \begin{tabular}{c c c c c }
        \toprule
        $d$ & $r$ & $\Pfpostest$ & $N_{\text{evals}}$ & $\mathrm{CoV}$ \\
        \midrule
        12  & 6 & $5.88\times10^{-3}$  & 311,000 & 0.017 \\
        30  & 12  & $1.491\times10^{-4}$ & 3,110,400 & 0.108 \\
        60  & 12 & $6.474\times10^{-4}$ & 6,220,800 & 0.057 \\
        150 & 16 & $6.13\times10^{-2}$ & 27,648,000 & 0.119 \\
        \bottomrule
    \end{tabular}
\end{table}     

Finally, to test the scalability of DIRT, posterior failure probabilities were estimated for increasing dimensions corresponding to larger numbers of KL coefficients in the random-field representations. Table~\ref{tab:pf_post_summary} summarizes posterior failure probability estimates across all dimensions considered. For \(d \in \{12,30,60,150\}\), DIRT produces stable estimates (maximum CoV around 12\%) across 3 independent runs under the chosen parameter values despite a substantial increase in stochastic dimension. For the highest-dimensional case \(d=150\), we adopt a maximum TT rank of \(r=16\). It is observed that DIRT is able to produce stable estimates of $\Pfpostest$ across all the dimensions considered in this work. It is also observed that the estimated values of $\Pfpostest$ do not converge with increasing dimensions. One possible reason for this behavior is that since additional KL modes introduce finer-scale spatial variability in the microscale fields which can either increase or decrease local stress concentrations leading non-monotonic changes in the failure probability. 

As expected, the total number of forward FEM evaluations required to build the transport maps increases with dimension from $311,000$ at $d=12$ to $27,648,000$ at $d=150$. Although this can seem computationally expensive, the result should be interpreted while keeping in mind the complexity of the problem. The stochastic dimension corresponds to KL coefficients of three random fields that jointly define the microscale material properties. These fields enter the homogenization process non-linearly and influence the macroscale stress response, ie., the failure event is governed by a high-dimensional and coupled mapping from microscale variability to structural response. This makes both posterior approximation and rare-event sampling significantly more difficult compared to problems with weakly coupled inputs. In this sense, the reported computational costs do not reflect the inefficiencies of the method but the inherent complexity of posterior reliability analysis in multiscale systems.   

An important observation is that the maximum TT rank $r$ grows only moderately with dimensions (from $r=6$ for $d=12$ to $r=16$ for $d=150$). This suggests that although the posterior and failure informed densities become more complicated in higher dimensions, they still admit useful low-rank structures that can be exploited by representing the densities in TT format.  


Overall, the numerical results confirm the effectiveness of the proposed framework across the full multiscale workflow. The Bayesian inversion step is able to recover the underlying microscale fields from limited noisy strain measurements with good accuracy, as reflected by the close agreement between the true and reconstructed fields. The VRNN surrogate then enables rapid and physically consistent evaluation of the homogenized stiffness tensors required in the macroscale forward model. Building on this, the DIRT-based sampling strategy yields stable posterior failure probability estimates with low variance even as the stochastic dimension increases, demonstrating the potential of the proposed approach for scalable multiscale structural reliability analysis under high-dimensional microscale uncertainty.

\section{Conclusions}
\label{sec:conclusion}
This work presented a multiscale structural reliability framework for fiber-reinforced composites with spatially varying microscale uncertainty. The central objective was to estimate posterior failure probabilities for multiscale problems while accounting for uncertainty in microscale RVE parameters represented as random fields. To enable repeated evaluations of the multiscale forward model at input realizations, we combined a physics-augmented homogenization surrogate with a transport-based importance sampling strategy. In particular, the Voigt--Reuss Neural Network was used to map uncertain microstructural parameters to homogenized stiffness tensors at macroscale integration points. The key advantage of this surrogate in a reliability context is physical admissibility: the predicted constitutive tensors are constrained within the Voigt--Reuss bounds, preventing nonphysical stiffness predictions that can occur with unconstrained regression models. This property is particularly important when the sampling method intentionally concentrates evaluations in low-probability regions of the input space, where extrapolation beyond the training distributions is more likely.

For posterior reliability estimation, we employed the Deep Inverse Rosenblatt Transport approach to construct a deterministic map that approximates the optimal importance sampling proposal for the failure and posterior distributions induced by strain measurements. In the low-dimensional setting ($d=12$), posterior failure probabilities obtained using the proposed approach are consistent with reference estimates from rejection sampling, providing verification of the posterior inference and reliability pipeline. Moreover, the methodology was demonstrated at stochastic dimensions up to $d=150$, illustrating practical scalability to random-field discretizations relevant to spatially heterogeneous composites.
Beyond estimator accuracy, the proposed approach offers practical advantages over subset-based reliability strategies such as Bayesian Updating with Structural Reliability - Subset Simulation \cite{BUS_varying} in the present setting. Subset simulation methods are effective for rare-event estimation, but they typically require Markov chain transitions at intermediate levels and careful tuning to maintain mixing in concentrated intermediate distributions, especially in high dimensions. In contrast, once a transport map is constructed, generating approximate posterior samples and evaluating failure probabilities can be performed without burn-in and can be parallelized across independent batches, which is advantageous in large-scale computational mechanics workflows. The deterministic map also offers an amortization benefit for repeated analyses under the same likelihood model and measurement configuration. In particular, if the map targets the posterior distribution, it can be reused to answer multiple reliability queries (e.g., different quantities of interest or sensitivity studies) by re-evaluating the corresponding failure indicators on posterior samples. When the map is tuned to a specific failure event, changes in the failure threshold may require additional refinement; nevertheless, the learned map provides a strong initialization that can reduce the marginal cost of subsequent analyses.

The main limitation observed is the growth in computational cost with dimension, driven by the high TT ranks required to represent increasingly complex posteriors and failure sets in TT format, together with the cost of each target function evaluation (a macroscale FE solve). Several directions can address this bottleneck. First, the ranks required to represent a function accurately in TT format depend strongly on the variable ordering \cite{cui2023scalable}; therefore, reordering and block grouping of KL coefficients can be investigated to reduce effective ranks and improve scalability. Second, one can consider additional surrogate modeling at the macroscale (e.g., polynomial chaos expansion \cite{voelsen2023pce} or NN surrogates \cite{maurizi2022gnn}) to decrease the cost per function evaluation. Third, we can construct the DIRT approximations to the optimal biasing densities in reduced subspaces. In particular, a DIRT approximation to the posterior can be constructed in a \emph{likelihood-informed subspace} \cite{zahm2022certified}, within which the posterior differs most significantly from the prior. When approximating $p^{*}$ or $p^{y*}$ (i.e., the optimal biasing densities containing the indicator function for the failure region), we can apply similar ideas to define a \emph{failure-informed subspace} \cite{uribe2021cross} within which the DIRT approximation can be constructed. Alternatively, we can apply the ideas from Refs.~\cite{tong2021extreme, tong2023large} which construct a suitable subspace using ideas from large deformation theory.

Additionally, we note that in this work, both the construction of the DIRT approximations to the optimal importance densities and the corresponding estimates of the failure probability use only the VRNN surrogate. Future work could investigate the use of a multifidelity estimation procedure \cite{peherstorfer2018survey} which uses a carefully-chosen combination of VRNN and FE$^{2}$ solves to obtain more accurate estimates of failure probabilities.


Finally, future studies will extend the present linear-elastic benchmark toward more realistic composite failure mechanisms, including progressive damage and fracture, where localized nonlinearities and path dependence introduce additional challenges.

\section*{Acknowledgements}
This material is based upon work partially supported by the U.S. National Science Foundation under award No. 2452029.  
The opinions, findings, and conclusions, or recommendations expressed are those of the authors and do not necessarily reflect the views of the NSF.
\\
The authors acknowledge the Texas Advanced Computing Center (TACC) at The University of Texas at Austin for providing computational resources that have contributed to the research results reported within this paper. URL: \url{http://www.tacc.utexas.edu}\\
De Beer and Cui are partially supported by the Australian Research Council grant FT250100199.

\appendix
\section{Validation of VRNN-Based Homogenization Against Classical FE$^2$}
\label{appendix:vrnn_validation}

This appendix presents a quantitative validation of the proposed VRNN surrogate model by comparison against the classical two-scale finite element homogenization FE$^2$ procedure. 

\begin{figure}[h!]
    \centering

\includegraphics[width=0.95\linewidth]{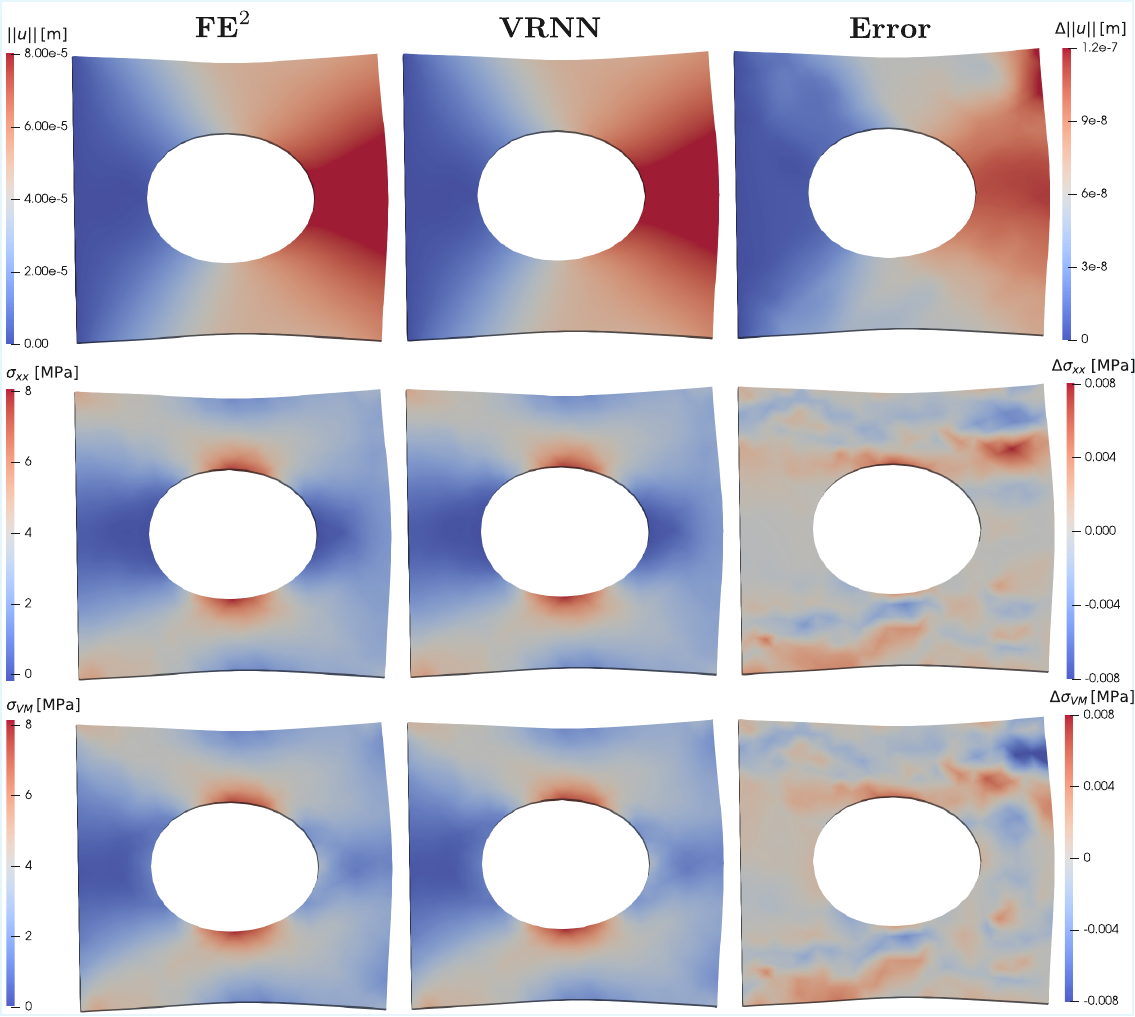}
    \caption{
        Comparison of the macroscale fields predicted using classical FE$^2$ homogenization and the VRNN surrogate on the deformed mesh (scaled by 200x).
        \textbf{Top:} $||u||$ displacement magnitude field.
        \textbf{Middle:} $\sigma_{\mathrm{xx}}$ stress field.
        \textbf{Bottom:} $\sigma_{\mathrm{VM}}$ von-Mises stress field.
        All three quantities show agreement between the two homogenization approaches. Strain fields and other stress component fields show similar agreements but are not plotted here. 
    }
    \label{fig:vrnn_validation_appendix}
\end{figure}

Figure~\ref{fig:vrnn_validation_appendix} illustrates a representative realization of the three microscale random fields considered in this study: the fiber volume fraction $v_f(x)$, the fiber Young’s modulus $E_f(x)$, and the matrix Young’s modulus $E_m(x)$. For this realization, the homogenized stiffness tensor $\mathbf{C}^\text{hom}(x)$ is computed at every material point using two approaches:
\begin{enumerate}
    \item the classical FE$^2$ method based on direct RVE simulations, and
    \item the VRNN surrogate model described in Section~2.4.
\end{enumerate}

The resulting elementwise stiffness tensors are subsequently used within the macroscale plate-with-hole problem, and a single forward finite element solve is performed for each homogenization strategy. By comparing the displacement, strain, and stress fields produced by both approaches, we evaluate the fidelity of the VRNN surrogate relative to the high-fidelity FE$^2$ predictions.

Overall, the close agreement between the FE$^2$ and VRNN-based solutions across all primary quantities of interest demonstrates that the VRNN surrogate is a reliable and computationally efficient alternative to classical two-scale homogenization for the considered class of random microstructural fields.

\clearpage

\bibliographystyle{unsrt}  
\bibliography{references}

\end{document}